\documentclass[10pt]{article}
\usepackage[letterpaper,margin=1in]{geometry}
\usepackage{iftex}
\ifPDFTeX
  \usepackage[T1]{fontenc}
  \usepackage[utf8]{inputenc}
  \usepackage{microtype}
\fi
\usepackage{lmodern}
\usepackage{amsmath,amssymb}
\usepackage{array,tabularx,booktabs}
\usepackage{enumitem}
\usepackage{fancyvrb}
\usepackage{xcolor}
\usepackage[colorlinks=true,linkcolor=blue!55!black,citecolor=blue!55!black,urlcolor=blue!55!black]{hyperref}
\setlist{itemsep=2pt,topsep=4pt}
\newcolumntype{Y}{>{\raggedright\arraybackslash}X}

\title{FlashProve-Next: Verifiable Inference for Production LLM APIs}
\author{Liang Song\\
\small Independent Researcher\\
\small \texttt{sl12160010@gmail.com}}
\date{November 2026}
\begin{document}
\maketitle
\begin{abstract}
Clients of closed-weight LLM APIs cannot verify whether a provider ran the advertised model, precision, hidden prompt, and reasoning budget. Existing approaches either trust hardware or third parties, or re-prove every layer of an integer surrogate at high cost while diverging from native floating-point serving.

FlashProve-Next produces verifiable, zero-knowledge-masked proofs for individual LLM API responses without modifying native serving (\texttt{bf16}/\texttt{fp8}/\texttt{mxfp4}, random sampling, FlashAttention, CUDA graphs, batching, prefix caching) and without disclosing weights, hidden prompts, hidden reasoning, or activations (dense profiles fix coarse geometry; cross-architecture hiding is deferred). A non-depleting Reed--Solomon weight commitment \texttt{model\_\allowbreak{}id} opens inside proofs via hidden column openings, while local per-operator rounding contracts remove integer surrogates and the refusal-bit leak. Cost-weighted spot-checking supports non-interactive public verification (\texttt{NI-315}, $\varepsilon \le 2^{-40}$, $Q \le 2^{64}$) and external-seed \texttt{C\_\allowbreak{}pre} mode (\texttt{ExtSeed-153}, $\varepsilon \le 2^{-20}$, $Q_{\rm samp}\le 2^{30}$).

Twenty-eight Lean 4 modules (696 declarations, zero \texttt{sorry}, standard axioms only) mechanize round-by-round soundness, bucketed sampling, $\mathbb F_{p^3}$ constrained-mask binding, causal masking, six IR gadgets, and executable \texttt{FPN2} verifiers. Five-model Blackwell calibration gives zero false rejections across 1.85M GEMM/MoE blocks and 1.96M non-GEMM rows. Our zero-Phase-2 folded backend (folded ZK simulator open) proves Qwen3-0.6B \texttt{ExtSeed-153} (1.47B cells, 8 MiB) in 450.2 s GPU (10.76 s CPU) at unique-decoding $Q=128$, passes a 14/14 cheating matrix, and scales to 36-layer Qwen3-4B (dev $4+2$) and live \texttt{vLLM} Qwen3-30B-A3B-FP8 Relation 6 (1.49 s prove, 23.4 ms Rust verify, public weight tiles). Rust \texttt{fpn-verify} checks 8 MiB \texttt{FPN2} Relation 2 proofs in 10.1 ms on CPU.

\end{abstract}
\section{Introduction}
\subsection{The problem}
Frontier model providers sell access to a named model through an API. What actually runs behind the endpoint is invisible to the client: a smaller or distilled model under load, an int4 or fp8 quantization of the advertised bf16 checkpoint, skipped layers or fewer active experts, a truncated context window, a cheaper system prompt or a reduced reasoning budget, unverified speculative drafts, or a cached answer to a similar earlier request. Each of these saves the provider money, and each is undetectable from the visible text alone at low substitution rates \cite{CSZS25,GLG25}.

The goal of FlashProve-Next is a statement of the following form, verifiable by anyone on an ordinary CPU from the published \texttt{model\_\allowbreak{}id}, the request, the visible response and a proof:

\begin{quote}The provider holds a computation trace consistent with the configuration committed in \texttt{model\_\allowbreak{}id} (weights, number formats, hidden prompts, sampling rule), the trace generated this visible response for this request under the committed sampling rule, including hidden prompts, injected content and hidden chain of thought, and the fraction of computation (by committed cost) outside floating-point tolerance and the fraction of generated tokens that violate the sampling rule are both below public thresholds.\end{quote}
The statement is about \emph{possession of a compliant trace}, not about which GPU executed what. Recomputing a compliant trace after the fact costs at least as much as honest inference, so it is not a way to save compute.

Three constraints make this hard simultaneously. \textbf{Closed weights and hidden topology}: the proof must not reveal weights, layer count, width, whether the model is a mixture of experts, its routing, or its deployed number format. \textbf{Private context}: hidden system prompts, tool templates and chain of thought are part of the computation and must stay hidden. \textbf{Unchanged production inference}: the commitment must be to what really runs, in native floating point with every acceleration, not to a fixed-point reference model invented for the proof.

\subsection{Why not trusted hardware, receipts or re-execution}
A TEE attestation \cite{CSZS25} moves the trust to the chip vendor's root keys, firmware and side-channel resistance, and ties the protocol to specific confidential-computing GPUs; it does not extend to TPUs or Trainium. A signed receipt is a statement by the party whose cheating we want to detect. Activation fingerprints and same-seed re-execution (TOPLOC \cite{ODP+25}, DiFR \cite{KRR+25}, FSSL \cite{RKH+25}, SVIP \cite{SLZJZ24}) are effective when the verifier holds the weights, the plaintext context and the sampling seed, which a closed-weight API client never does. Refereed delegation (Verde \cite{AST+25}) needs at least one honest replica, and proof-of-sampling schemes \cite{ZWT+24,WZS+25} need staked verifiers who hold the model. None of these fits a single closed provider facing the public.

\subsection{Why not full zkML}
Full zero-knowledge proofs of inference (ZKML \cite{CWSK24}, zkLLM \cite{SLZ24}, zkGPT \cite{QSL+25}, FlashProve \cite{RAC+26}) satisfy the trust requirement but prove a fixed-point or integer surrogate of the model. This has two consequences. First, cost: FlashProve, the most efficient system we know of, reports a per-GPU proving time around $71\text{--}93\times$ the prefill time of an 8$\times$H100 FP8 vLLM deployment \cite{RAC+26}. Second, and more fundamental, the surrogate is not what runs. A native bf16 or fp8 engine diverges from the integer model on long sequences and near sampling ties, and an honest provider then cannot produce a proof. Under strict zero knowledge, \emph{whether a proof can be produced} is itself a bit that depends on hidden activations; Section 10 shows that any per-response numerical refusal rate above zero gives a distinguisher with constant advantage after enough queries. Sampled variants (SLP \cite{LY26}) cut cost but keep the surrogate and, in the prototype, drop zero knowledge.

\subsection{Approach}
FlashProve-Next decouples verification from online serving. The provider publishes a one-time hiding commitment \texttt{model\_\allowbreak{}id} to its real deployed weights, numerical formats, hidden prompts, sampling key, and per-operator rounding contracts, accompanied by a one-time zero-knowledge setup proof $\pi_{\rm setup}$. During online serving, the inference engine runs unmodified in native floating point while lightweight read-only hooks record boundary tensors; after the response finishes, a background prover commits the trace and proves correctness of only $k$ cost-weighted random samples of computation blocks and generation steps chosen by Fiat--Shamir. Each sampled block is checked against a \emph{local} single-hop reference evaluated inside the proof on the engine's own committed inputs, paired with a box-plus-ball error radius propagated from the committed per-operator rounding contract, while weights are authenticated via in-circuit hidden column openings with zero static-codeword wear-out across unbounded proofs. Because reference values never chain across blocks, error radii remain at single-operator rounding scale, and honest native-FP execution is \emph{provably} accepted whenever deployed kernels satisfy their rounding contract. Any cheat that saves substantial compute must either emit out-of-radius outputs on skipped or quantized blocks---caught with probability $1-(1-\sigma)^k$ per attempt---or pay the honest compute cost.

\subsection{Contributions}
\begin{enumerate}
\item{} \textbf{A versioned API response relation} (Section 4; \texttt{Spec 01}, \texttt{Spec 05}) covering canonical request encoding, typed segmented tokenization, hidden prompt injection, zero-freedom object wiring, stratified cost-weighted sublayer lineages, constrained decoding, byte-level Aho--Corasick stop semantics, tool calls, and acyclic multi-turn encrypted state anchors.
\item{} \textbf{A non-exhausting static model commitment} (Section 5; \texttt{Spec 09}) that commits deployed weights and radius matrices once via Reed--Solomon column trees and authenticates sampled GEMMs through two-round tile reductions and in-circuit hidden column openings (\texttt{HCO}), achieving unbounded lifetime zero knowledge, prover work proportional to touched tiles, a single-transcript setup proof $\pi_{\rm setup}$ requiring no correlated-agreement assumption, and compact in-circuit opening of hidden architecture and prompt metadata ($\approx 4\times 10^5$ data cells per proof).
\item{} \textbf{Certified per-operator rounding contracts and lineage cost accounting} (Section 6; \texttt{Spec 06}) establishing forward box-plus-ball propagation templates (including online-softmax FlashAttention with additive exponential underflow), per-GEMM observation granularity, and $h=3$ consumer-lineage billing (\texttt{Rule CA}) that closes the zero-operand trivialization loophole.
\item{} \textbf{Sort-free integer bucketed keyed sampling and zero-blowup constrained decoding} (Sections 4.5, 7; \texttt{Spec 05}, \texttt{Spec 06}) replacing vocabulary-wide sorting and per-token PRF calls with 32-token buckets and a witness-free per-token nucleus sandwich (\texttt{sure}/\texttt{poss}), paired with an $\mathbb F_{p^3}$ multilinear Schwartz--Zippel binder that enforces compiled JSON-Schema byte-DFA vocabulary masks in $O(k_2 Q_{\rm pad})$ extra cells without per-token vocabulary transitions.
\item{} \textbf{Lifetime knowledge soundness, adaptive multi-proof zero knowledge, and three-part completeness} (Sections 8--10; \texttt{Specs 02--04}), together with a \textbf{per-request KV matrix commitment and hidden-selection sumcheck} (\texttt{Spec 10}) that eliminates online GPU hashing and shrinks the long-context attention circuit slope by more than two orders of magnitude.
\item{} \textbf{Lean 4 formalization, folded \& streaming $\mathbb F_{p^3}$ provers, zero-dependency Rust verifier, and live \texttt{vLLM 0.\allowbreak{}33} evaluation} (Sections 12, 14; \texttt{Spec 07}, \texttt{Spec 11}): 28 Lean 4 modules (696 axiom-audited declarations, zero \texttt{sorry}, zero \texttt{native\_\allowbreak{}decide}), hardware calibration across 392,928 Ada blocks plus 1.85M GEMM/MoE/router blocks and 1.96M non-GEMM rows on five Blackwell checkpoints (\texttt{0} honest false rejections), a streaming Stage 1 prover (\texttt{1.\allowbreak{}51 B/cell}, 8.67G-cell $\pi_{\rm setup}$ and 5.25G-cell \texttt{NI-315} proof on \texttt{Qwen3-0.\allowbreak{}6B}), a v1.0 zero-Phase-2 folded backend (\texttt{8 MiB} \texttt{FPNZK2} on \texttt{Qwen3-0.\allowbreak{}6B} \texttt{ExtSeed-153} and 36-layer \texttt{Qwen3-4B}, plus zero-sync async capture and live \texttt{vLLM} \texttt{Qwen3-30B-A3B-FP8} Relation 6), and a zero-dependency \texttt{\#![forbid(unsafe\_\allowbreak{}code)]} Rust \texttt{fpn-verify} binary with detached pure-Python reference for \texttt{FPN2} Relations 1--6.
\end{enumerate}

\subsection{Evidence tags, artifacts, and hardware baseline}
All companion specifications (\texttt{specs/01}--\texttt{11}), the 28-module Lean 4 formalization (\texttt{lean/}), the zero-dependency Rust \texttt{fpn-verify} binary (\texttt{core/}), the detached Python reference verifier (\texttt{verifier/python/}), the prover and \texttt{vLLM 0.\allowbreak{}33} integrations, and all raw GPU benchmark JSON receipts are publicly available at \url{https://github.com/s-JoL/FlashProve-Next} (project page: \url{https://s-jol.github.io/FlashProve-Next/}). Every technical claim carries an explicit evidence tag:

\begin{center}\small
\begin{tabularx}{\linewidth}{>{\hsize=0.55\hsize}Y>{\hsize=1.45\hsize}Y}
\toprule
{}\textbf{Tag} & {}\textbf{Meaning} \\
\midrule
{}[Def] & {}protocol definition or design choice \\
{}[Paper] & {}mathematical proof in this paper / companion specifications \\
{}[Lean: \texttt{name}] & {}machine-checked statement in Lean 4.29.1 + Mathlib (\texttt{lean/THEOREM.\allowbreak{}md}) \\
{}[Assume] & {}explicit cryptographic, kernel, or systems assumption \\
{}[Meas: model, machine] & {}empirical measurement on the named checkpoint and hardware \\
{}[Est] & {}analytical operation/bandwidth calculation from measured slopes \\
{}[Open] & {}open question deferred to Stage 2 / future work \\
\bottomrule
\end{tabularx}
\end{center}

Unless noted otherwise, Stage 1 GPU experiments ran on an NVIDIA RTX 4070 Ti SUPER 16 GB (Ada Lovelace, \texttt{sm\_\allowbreak{}89}) with an Intel i7-14700F CPU and 25 GB RAM under Linux (\texttt{torch 2.\allowbreak{}14.\allowbreak{}0+cu130}, \texttt{transformers 5.\allowbreak{}20.\allowbreak{}2}), abbreviated \textbf{4070TiS} below; v1.0 folded-backend, Blackwell FP8/MXFP4 calibration, and async \texttt{vLLM} capture benchmarks ran on an NVIDIA RTX 6000D 84 GB (Blackwell, \texttt{sm\_\allowbreak{}120}, 120 GiB RAM), abbreviated \textbf{RTX 6000D}; lightweight CPU unit, Rust \texttt{fpn-verify}, and synthetic-layer tests ran on an Apple M1 Mac.

\section{Threat Model and Goals}
\subsection{Parties}
The \textbf{provider} runs inference and produces proofs. For soundness it is fully malicious: it holds all weights, keys and user data, chooses configurations (including malicious ones), requests to answer, salts, commitments and submission order, may grind, abandon and retry proofs, and may collude with tenants. Neither the provider nor any hardware vendor is trusted. \textbf{Tenants} send requests and receive responses and proofs; proofs may be forwarded, so zero knowledge must hold for any holder. \textbf{Verifiers} are anyone with \texttt{model\_\allowbreak{}id}, the request, the visible response and the proof.

\subsection{The provable layer}
\textbf{Security claim [Def].} For any adaptive, grinding provider, over the whole service lifetime and at most $Q$ first queries to the challenge oracle, the probability that \emph{any} accepted response has (a) bad blocks of cost fraction at least $\sigma$ or (b) generated tokens violating the committed sampling rule at fraction at least $\rho$ is at most $\varepsilon$. Defaults: $Q=2^{64}$ (covering re-salting, re-committing, abandoned and retried proofs), $\varepsilon\le2^{-40}$; $\sigma,\rho$ public. Cost weights are fixed by \texttt{model\_\allowbreak{}id}, never declared by the prover. The required notion is \emph{knowledge} soundness, because the claim says the provider \emph{holds} a trace (Section 8.1).

\textbf{What the thresholds mean.} The relation tolerates bad computation below $\sigma$ and sampling violations below $\rho$ with probability 1: at the default $\sigma=\rho=0.5$ a provider may skip up to just under half of the committed cost in a stratum, or substitute just under half of the generated tokens, on every response, and still be accepted. The defaults are chosen against the realistic cheat catalog below, whose members are coarse-grained: per-request routing to a cheaper model makes every block of that response bad ($d_C\approx1$), quantizing all GEMMs makes every quantized GEMM block bad (measured $d_C=0.681$ on raw per-GEMM blocks, rising to $d_C=1.000$ under $h=3$ rule CA lineage billing; Section 12.4), and layer skipping is charged at least its saving under rule CA. Fine-grained mixing below $\sigma$ (for example, substituting 49\% of tokens of every response) requires both models resident and saves at most $\sigma$ of the cost; it is inside the guarantee gap by definition. Both thresholds are public and configurable per profile, and the cost is explicit (Section 8.5): with $Q=2^{64}$, $\varepsilon=2^{-40}$ the three pools need $3\times105=315$ samples at 0.5, $3\times205=615$ at 0.3, $3\times327=981$ at 0.2 and $3\times691=2{,}073$ at 0.1, and proving cost grows roughly linearly in the sample count ($1.95\times$, $3.1\times$, $6.6\times$).

\subsection{The empirical layer: a realistic cheat catalog}
The provable layer bounds the fraction of bad computation; whether a given money-saving cheat produces bad computation is an empirical question, answered per \texttt{model\_\allowbreak{}id} by measurement and published with it. We restrict attention to cheats that are common in practice when the provider holds the full weights:

\begin{enumerate}
\item{} \textbf{Model substitution}: smaller, distilled or older model; routing to a cheaper model under load.
\item{} \textbf{Quantization}: int4/fp8 weights, KV-cache quantization, reduced-precision attention.
\item{} \textbf{Pruning-type}: structured pruning, layer skipping, fewer MoE experts, low-rank approximation.
\item{} \textbf{Context truncation}: sliding windows, KV eviction, dropping earlier turns.
\item{} \textbf{Hidden-configuration substitution}: different system prompt or template, lower reasoning effort.
\item{} \textbf{Unverified speculative decoding}: accepting draft tokens without target verification.
\item{} \textbf{Replay}: returning a cached response to an identical or similar request.
\end{enumerate}

Classes 5 and 7 are blocked by the commitment and chain checks directly (hidden configuration is committed; client nonces prevent replay), independent of sampling. For the others we measure ``fraction of compute saved $\to$ bad-cost fraction''. Two further requirements: the compute that can be saved \emph{within} tolerance for each \texttt{model\_\allowbreak{}id} must be strictly below $\sigma$ (target $\le10\%$; long-context attention tails are excluded from this target by decision, item (5) of Section 2.4), and ``laundering'' (generate with a cheap model, then prefill the real one) must map every non-compliant final token to $\rho$; if all final tokens comply, laundering is legitimate speculative sampling.

\subsection{Explicit non-goals}
FlashProve-Next does not protect against: (1) small tampering with output tokens or nudging of sampling after full compute was paid (fraction below $\rho$ is not guaranteed to be caught), including the provider grinding its own server nonce; (2) the model itself strategically shortening its chain of thought under the committed prompt; (3) the truth of external tool outputs; (4) \emph{exotic cheats engineered against the proof system's internal checks}, such as zeroing a causal prefix at one layer so later layers are formally consistent (``prefix collapse''), or feeding all-zero operands so that downstream matrix products are trivially exact (``zero-operand padding''); cheap mitigations are kept, expensive ones are not adopted; (5) \emph{sparse-attention savings whose output difference is below rounding scale}: dropping attention keys with very low weight. Any tolerance that does not reject honest execution must admit rounding at the precision in which attention probabilities are stored, and such drops perturb outputs by the same amount. We measured the oracle upper bound (Section 12.5) and require each \texttt{model\_\allowbreak{}id} to publish it per context length.

We state the exact magnitudes of both scope boundaries explicitly rather than obscuring them. \emph{Prefix collapse} can save a lot: zeroing the prefix's residual stream above layer $\ell_0$ removes a fraction $(1-\ell_0/L)$ of the prefix's computation (92\% for $\ell_0=6$, $L=80$; about 70\% of total prefill and KV compute), while only the lineages that reach layer $\ell_0$ within depth $h$ are charged, so $d^{(3)}_C$ can stay near $0.04$ [Est, consistent with the finite-lineage bound of Spec 06 Prop. 6.6]. Its effect on output quality is unmeasured. It is not the same as the layer-wise KV eviction used in practice (LazyLLM, PyramidKV): those drop keys from the softmax, so the attention outputs of the affected blocks leave the certified radius unless the dropped mass is below rounding (the tail-dropping case), and every key atom of such a block is charged through rule CA. To stay locally consistent a cheater must instead publish zero KV objects \emph{and} keep their phantom softmax weight $e^{-m}$ per zero key, a computation that exists only to satisfy FlashProve-Next's checks; this is why it is classed as exotic. \emph{Tail dropping} is measured: with an oracle on Qwen3-4B (4070TiS), 12--15\% of compute at 4k and 34--55\% at 16k--32k when only outputs are observed, 6--8\% and 19--37\% when the log-sum-exp is also observed (Section 12.5). Also out of scope for version 1: logprobs, user-chosen sampling parameters, and content produced by \emph{other} models (reasoning summaries, external classifiers, compaction), which can be covered by committing a separate \texttt{model\_\allowbreak{}id}.

\subsection{Zero knowledge and completeness requirements}
The simulator may see only: what the standard API already returns (request, visible response, usage counts including reasoning-token counts, cache-hit counts, service tier), \texttt{model\_\allowbreak{}id} and its update events, a public profile of size upper bounds, and two secret-independent protocol objects: the public service policy (refusal rules, deadlines) and a public outage log. Protected: weight values, depth, width, attention form, MoE structure and routing, number formats, hidden prompts, hidden chain of thought, activations, logits. Proof size, arrival time, queueing and failure must not depend on secrets; batching must not expose co-batched requests; and the bit \emph{whether a proof can be produced} must not leak. Completeness target: an honest service fails to produce a proof with probability at most $2^{-20}$ per response, including adversarially constructed prompts.

\section{System Overview}
\begin{center}
\begin{Verbatim}[fontsize=\fontsize{6.2}{7.2}\selectfont]
 publish once                         per request (online)                    per request (background)
 ------------                         --------------------                    ------------------------
 weights W_raw (bf16/fp8/...)         client: x_pub, nonce N_c                 prover worker
   | canonical decode CD                 |                                       |
   v                                     v                                       v
 RS-encode rows, column Merkle    engine runs as usual (native FP,       open committed objects at
 tree  -> C_W, C_raw              FlashAttention, batching, TP/EP,       Fiat-Shamir sampled blocks
 stable tables (graph, contract,  speculative decoding, caching)         and generation steps
 tiles, costs) -> config rows of W    | read-only hooks                     |
 hidden prompts, K_prf -> C_cfg       v                                     v
   |                              boundary tensors -> salted sponge      certified local reference
   v                              leaves -> trace root R; global         + radius per sampled block;
 model_id = H(salt, C_W, C_raw,   columns (tokens, slot table,           sampling relation per step;
             C_cfg) + pi_setup    automata) -> cm_glob                   HCO weight opening; ZK proof
   |                                  |                                     |
   +------- official location         +--> response y, N_s, proof_id,       +--> fixed-size proof at a
                                           R, state anchor (stream end)          fixed release time
                                                                                 |
                       verifier (any CPU): model_id + x_pub + y + proof --> accept / reject
\end{Verbatim}
\end{center}
\emph{Figure 1. Data flow. The online path adds read-only observation hooks and hashing; proof generation runs in the background and never blocks inference.}

\section{The Relation}
\subsection{Configuration, statement and witness}
The model configuration is [Def]

\[M=(W,\mathsf{dtype},\mathsf{graph},\mathsf{Tok},\mathsf{Pre},\mathsf{Render},\mathsf{Inject},\mathsf{Sample},K_{\rm prf},\mathsf{MaskCompile},\mathsf{Stop},\mathsf{Cost},\mathsf{Tol},\mathsf{Caps}):\]

weights as raw deployed bit strings with scales and interpretation rules, the canonical logical graph, tokenizer and preprocessing, rendering templates and hidden injections, the sampling rule and its key, the schema-to-mask compiler, stop semantics, the committed cost table, the rounding contract, and the capability matrix. Everything except the weights is small and is opened fully in every proof; the weights are committed once, as described in Section 5, and the relation's weights are \emph{defined} as the unique decoding $W_*$ of that commitment, fixed at publication time. There is therefore no per-block ``missing weight'' case.

The public statement is

\[X_0=(v_{\rm rel},\mathsf{model\_id},x_{\rm pub},y,\ell_{\rm api},\mathsf{profile},N_c,N_s,e,\mathsf{sess},\mathsf{St}_{\rm prev},\mathsf{St}_{\rm new},\sigma,\rho):\]

protocol version, commitment, public request, response, the usage fields the API already returns, profile, client and server nonces, key epoch, session identifier and state anchors, thresholds. The cryptographic statement adds the trace root $R$ and the commitment $\mathsf{cm}_{\rm glob}$ to the global witness columns. The witness contains the configuration opening, hidden injections, hidden reasoning tokens, tool returns, the full logical token sequence $u$, the object table, the private slot table, automaton state columns, and the openings of the state anchors.

A response is split into a core payload and an envelope, $y=(y^{\rm core},\mathsf{env})$ [Def]. The core is the canonical concatenation of API content blocks, finish reason, usage and stop echo; the envelope holds the fields the proof system adds ($N_s$, new state anchor, encrypted state blob, \texttt{proof\_\allowbreak{}id}). This split is what makes the multi-turn chain acyclic (Section 4.6).

Request canonicalization is an injective, typed, length-prefixed encoding that keeps exact Unicode, distinguishes absent, null, empty string and empty list, keeps semantically ordered arrays, and binds media by content bytes and preprocessing version. Rendering produces typed segments (control, text, multimodal placeholder, tool return, replayed hidden history) that are tokenized separately; control tokens can only come from control segments and merges never cross segment boundaries [Def]. An assembled token vector is never an unchecked free witness.

\subsection{Object table and zero-freedom wiring}
Every cross-block tensor is an immutable \emph{object}: the actual matrix-multiply and attention operands (after quantization, as code plus scale), block outputs, residual stream values, newly produced carried state (KV entries, MLA latents, recurrent state versions), and the final normalized hidden state $x^{\rm fn}$ [Def]. Logits are \emph{not} objects: they are never written out or hashed online; for a sampled generation step the circuit recomputes reference logits from the committed $x^{\rm fn}$ and the committed LM-head weights (Sections 4.5 and 5). An object key $\kappa=(\mathsf{seg},t,s,\mathsf{port},\mathsf{ver})$ maps under a public injective function $\mathsf{Addr}$ to an address whose range depends only on the profile. The trace commitment is a position-binding Merkle tree

\[R=\mathsf{MT}_{H_{\rm in}}\big((\ell_a)_{a}\big),\qquad \ell_a=H_{\rm in}\big(\texttt{dom}_{\rm leaf},a,\mathsf{salt}_a,\mathsf{Enc}(v_a)\big),\]

where each leaf absorbs the \emph{complete} value with a fresh 256-bit salt. The published root is salted once more, $R=H_{\rm in}(\texttt{dom}_{\rm root},\varsigma_R,\mathsf{MT}(\cdot))$ with a fresh 256-bit $\varsigma_R$ per request: without it, a legal empty response (Section 4.5), whose tree holds only fixed padding leaves, would have a deterministic root while the simulator outputs a uniform string. A block's input addresses are computed inside the proof from the committed graph and slot table: if the graph connects output port of slot $s_1$ to input port of slot $s_2$, then $\mathsf{Addr}(\mathrm{in}_{s_2})=\mathsf{Addr}(\kappa_{\mathrm{out}_{s_1}})$. Producer and every consumer read the same leaf, so the wiring leaves the prover no freedom [Paper, Spec 05 Lemma 1.3]. KV entries are written once; recurrent state gets a new version per chunk. No format conversion is allowed on an edge: casts and quantization are the tail of the producing block, so the observed object is exactly the next operator's operand.

Leaves must hash complete values with a hash the extractor can observe. An earlier design used compressing algebraic digests ($D=Av+Br$, Byte-SIS). Such digests have astronomically many short preimages, so ``a consistent trace exists'' is trivially true, and computing them does not query an oracle, so a straight-line extractor cannot see unsampled blocks; rewinding extraction needs on the order of $Q/\varepsilon$ runs [Paper]. Byte-SIS can remain only as an acceleration layer that carries no soundness.

\subsection{Blocks, certified reference and bad-block indicator}
For a valid slot $b$ with inputs $(a_b,q^{\rm in}_b)$ (operands and read carried state) and outputs $(o_b,q^{\rm out}_b)$, the certified reference function is

\[\big(\hat o_b,\hat q^{\rm out}_b,\mathbf r_b,\varrho_b\big)=F^\sharp_{M,b}\big(a_b,q^{\rm in}_b\big),\]

one deterministic, integer-encoded hop from the engine's \emph{committed} inputs that also propagates the rounding contract through the block's operator DAG to a box radius $\mathbf r_b$ and a ball radius $\varrho_b$ [Def]. It is not a fixed-point reference model and never chains across blocks. Tolerance is a predicate of inputs and outputs:

\[\mathrm{Tol}_{M,b}\iff \sum_i\big(|\hat v_i-v_i|-r_i\big)_+^2\le\varrho^2\ \wedge\ \mathrm{Out}(b)\in\mathrm{Cap}_b\ \wedge\ \mathrm{Fin}\ \wedge\ \mathrm{In}(b)\in\mathrm{Rng}_b,\]

a box-plus-ball test, an output-range constraint, finiteness, and an input-range condition that guarantees no wrap-around in the field embedding. Only recurrent state objects may have $\varrho>0$.

The bad-block indicator $V_b$ is 1 if any object read or written by $b$ has no known preimage (written $\bot$ by the extractor), is non-canonically encoded or has the wrong type or shape; otherwise $V_b=\mathbf 1[\neg\mathrm{Tol}_{M,b}]$ [Def]. Weights enter a block only as hidden tile claims bound to the weight commitment by the hidden column opening of Section 5. The head block has no output object: its certified reference \emph{is} the reference logits with their radius, so $V_{\rm HEAD}$ checks only its input, and any deviation in the engine's LM head shows up in the sampling predicate. A missing or illegal \emph{local} object makes only the blocks that reference it bad, at their full cost; it never falsifies the global predicate. This \emph{locality} is what lets the soundness proof argue that accepting a sampled bad block requires inverting the hash.

\subsection{Costs, strata and the threshold relation}
Costs $c_b\in\mathbb N_{>0}$ come from the committed cost table by graph node [Def]. Every block that must execute appears exactly once, and the prover cannot reprice, drop bad blocks, duplicate good ones or assign zero weight. The private slot table (validity, position, node, expert, implementation bit, stratum, cost, prefix sums) is committed before any sampling challenge, and a set of global constraints forces it to be the unique unrolling of the committed graph on the request's data (token count, segments, multimodal patch count, MoE routes, implementation bits) [Paper, Spec 05 Lemma 1.1]. Prefix-cache hits, replayed hidden history and in-turn tool returns stay in the pool at full weight; there is no exemption for content ``already covered by an earlier proof''. Unaccepted speculative drafts are not in the pool and cannot dilute it.

A single cost pool has a dilution problem: with a 10,000-token prompt and 500 generated tokens, generation columns are about 5\% of cost, so computing all of them with a cheap model gives $d_C\approx0.05<\sigma$, while decode is where cheating saves the most money. A public stratification rule $\Pi$ therefore splits blocks into strata, by default $\Pi_2$: columns whose logits are sampled (generation stratum) versus all others (input stratum) [Def].

The threshold is counted on \emph{cost atoms over a lineage}, for the reasons given in Section 6.4 [Def]. Each block's cost is split by a public rule into atoms (one self atom, plus one key atom per key read by an attention block); operators are grouped into accounting units (attention sublayer, MLP sublayer, routed expert, router+combine, head+sampling) whose check set is every block containing one of their operators; the lineage $\mathrm{Lin}_h(a)$ of an atom is its unit plus the units below it along public dependency edges up to depth $h$ (default 3), and for a key atom also the producing column of that key. An atom is bad, $V_a=1$, if any block in the check set of any unit of its lineage is bad, and $d^{(j)}_C=\sum_{a\in A_j}c_aV_a/C_j$. The locality condition becomes: $V_a$ depends only on global data and the objects referenced by the blocks of its lineage, all of which are opened in the circuit when $a$ is sampled. With these per-stratum fractions and generation violation fraction $d_G$, the relation is

\[\mathcal R_{\Pi,\sigma,\rho}(X,w)= E\wedge\mathrm{Chain}\wedge\max_j d^{(j)}_C<\sigma\wedge d_G<\rho,\]

where $E$ checks canonical encoding of global objects and $\mathrm{Chain}$ the global consistency conditions below. Since $V_a$ is at least the indicator of the atom's own block and $d_C$ is a convex combination of the $d^{(j)}_C$, this implies the single-pool block-level condition of our goals; the relation only gets stricter, and honest executions have no bad blocks.

\subsection{Generation, constrained decoding and stop semantics}
Generation steps are numbered in logical order across hidden reasoning, visible text, generated tool arguments, refusals and end tokens, including tokens cut from the visible projection by a stop sequence [Def]. The sampling predicate $S_t$ reads reference logits and per-component certified radii that the head block recomputes inside the circuit from the committed $x^{\rm fn}$ at the step's position, the allowed set $A_t$, and the keyed noise (Section 7); no logits are committed and there is no separate ``sampling logits''. The engine samples from its own fp32 logits, which lie within the radius under H-contract. Constrained decoding is enforced \emph{exactly}, not within $\rho$: $\mathrm{Chain}$ checks $u_t\in A_t$ and the transition of a minimal co-accessible byte-level DFA ($|Q|\le Q_{\max}=128$) compiled deterministically from the public schema at every step. Rather than running $V$ byte-DFA transitions per sampled step inside the circuit, each sampled step $s$ commits a boolean allowed-mask column $a_s(v)$ alongside a one-hot state indicator $e_q(s,\cdot)$ over $Q_{\rm pad}$; after Phase 1, a fresh Fiat--Shamir point $\mathbf r_v\in\mathbb E^{\lceil\log_2 V\rceil}$ checks $\sum_{v<V}\mathrm{eq}(\mathbf r_v,v)\,a_s(v)=(1-\mathrm{act}_s)\sum_{v<V}\mathrm{eq}(\mathbf r_v,v)+\mathrm{act}_s\sum_{q<Q_{\rm pad}}e_q(s,q)\,m_q(\mathbf r_v)$, where both prover and verifier evaluate the $Q_{\rm pad}$ public scalars $m_q(\mathbf r_v)=\sum_{v\in A(q)}\mathrm{eq}(\mathbf r_v,v)$ outside the circuit in milliseconds [Def; Paper; Lean: \texttt{ConstrainedMask}]. Over $\mathbb E=\mathbb F_{p^3}$, any single-token mask discrepancy fails with probability at least $1-\lceil\log_2 V\rceil/\lvert\mathbb E\rvert>1-2^{-186}$, while adding only $O(k_2 Q_{\rm pad})$ (\textasciitilde{}$8\times 10^3$) committed cells. The supported schema dialect is fixed per version (types, required and bounded optional properties, bounded arrays and strings, small integer ranges, \texttt{enum}/\texttt{const}, \texttt{anyOf}, bounded-depth \texttt{\$ref}); unsupported keywords such as \texttt{uniqueItems} are rejected by a public rule, which leaks nothing [Def]. Compilation trims non-co-reachable states so every legal prefix extends to a complete legal output.

Stop sequences in the full relation use byte-level first-hit semantics: on the canonical byte stream of stop-eligible content, the first end position at which any stop string matches, with the longest match at that position; an Aho--Corasick automaton covers hits inside a token, across tokens and overlapping strings [Def]. (In the Stage 1 \texttt{vLLM} implementation, endogenous stop tokens---\texttt{EOS} and \texttt{</tool\_\allowbreak{}call>}---and \texttt{max\_\allowbreak{}tokens} budgets are proved in-circuit, while arbitrary user-supplied string \texttt{stop} arrays are excluded deterministically by the public pre-check $\Phi_{\rm pre}$.) Events in one step follow a fixed total order: pre-sampling budget guard (including \texttt{max\_\allowbreak{}tokens=0}), then endogenous control tokens, then any stop hit. A response with zero generated tokens is legal only in three public branches: the budget guard fires at step 0, a public policy code says an external classifier blocked the request before the main model ran, or the client cancelled before the first token.

\subsection{Nonces and the multi-turn state chain}
The client chooses a fresh nonce $N_c$ of at least 128 bits per request, which prevents replaying an old response with its proof; without it the provider could reuse the server nonce and replay [Paper]. The server draws a fresh 256-bit $N_s$ per attempt from a CSPRNG and returns it in the first stream event; no counter is used, because a monotone counter visible to clients leaks other tenants' request volume. $N_s$ makes sampling-noise inputs distinct with overwhelming probability regardless of how clients choose $N_c$, which blocks same-nonce probing of logits.

Multi-turn reasoning APIs return encrypted reasoning state that the client sends back. FlashProve-Next binds turns through a public hiding anchor

\[\mathsf{St}_i=H_{\rm in}\big(\texttt{dom}_{\rm st},v_{\rm rel},\mathsf{model\_id}_i,\mathsf{sess},i,\mathsf{St}_{i-1},H(x^{\rm core}_i),H(y^{\rm core}_i),H(\mathsf{hist}_i),r_i\big),\]

where $\mathsf{hist}_i$ is the hidden history handed to later turns, computed in the order $y^{\rm core}_i\to\mathsf{hist}_i\to\mathsf{St}_i\to\mathsf{blob}_i=\mathsf{AEAD}(\mathsf{aad}_i,(\mathsf{hist}_i,r_i))\to\mathsf{env}_i$, so the anchor commits only the core payload and the commitment graph is acyclic [Def]. The next turn opens $\mathsf{St}_{\rm prev}$ and checks session, turn index, history digests and the exact rendering of the replayed hidden history. Validity of a whole conversation follows by finite external induction over the public anchor chain; no proof is verified inside another proof, so no in-circuit random oracle is needed. Rollback and fork detection require the client to keep its latest verified anchor [Def].

\section{Model Commitment}
\subsection{Two constraints that shape the design}
\textbf{Static codeword exhaustion [Paper, Spec 09 Lemma 2.1].} For any hash-based polynomial commitment (Basefold, FRI, WHIR, Ligero) that opens the same static codeword in \emph{public} transcripts, the number of revealed positions per row grows with the number of openings; once it reaches message length plus blinding, MDS decoding recovers the row. Per-opening fresh masks do not help, because the verifier must read the static symbols. Lifetime zero knowledge with unbounded openings therefore requires that static leaves are never public, or a homomorphically re-randomizable commitment. \textbf{Protected positions [Paper].} Which matrix of which layer was sampled is itself protected; evaluating at a public point indexed by the matrix reveals it.

\subsection{Construction: static RS commitment with hidden column openings}
\textbf{Commitment [Def].} Weights are arranged by public tile shape $T_o\times T_i$ into a matrix $\mathbf W\in\mathbb F_p^{n_r\times n_c}$, one tile per row, $n_r=R_{\max}$ from the profile. The lower half of the rows holds the canonical signed-exponent block-floating-point decoding $\mathbf W^{\rm cd}$ used in proofs: a scale row (the decode exponent of every row), a row of per-tile-row squared norms, vector rows (normalization gains), configuration rows (Section 5.3) and the tiles. Each row has one decode exponent $G\in[12,28)$, required to be the \emph{largest} that keeps the row's integers below $2^{22}$, so a provider cannot quantize a row by under-scaling it. The upper half holds the \emph{radius matrix} $\mathbf W^r=\lvert\mathbf W^{\rm cd}\rvert+\kappa_n m$ at row $R_{\max}/2+t$ for tile row $t$, where $m$ flags the entries whose decoding is inexact and $\kappa_n$ depends on the contract and the input width: certified tolerance needs $\lvert W\rvert\lvert x\rvert$ (Section 6.2), which is not linear in $\mathbf W^{\rm cd}$, so it is committed as a second half of the same matrix and opened by the same mechanism. A second matrix $\mathbf W^{\rm raw}$ holds the deployed bit strings, used only for binding to what actually runs. Each row is Reed--Solomon encoded at rate $\vartheta=1/4$, each column gets a 256-bit salt, and a Poseidon2 ($H_{\rm in}$) Merkle tree over columns gives $\mathsf C_W$ (resp. $\mathsf C_{\rm raw}$). Then

\[\mathsf{model\_id}=H_{\rm in}\big(\texttt{dom}_{\rm mid},\varsigma_0,\mathsf C_W,\mathsf C_{\rm raw},\mathsf C_{\rm cfg}\big),\]

where $\mathsf C_{\rm cfg}$ commits only the volatile configuration (sampling key, hidden prompts and templates; Section 5.3).

\textbf{Setup proof [Def; Paper].} $\pi_{\rm setup}$ is published once with \texttt{model\_\allowbreak{}id}. It proves in zero knowledge that (i) every row of both codeword matrices agrees with the RS encoding of a \emph{committed message row} on one common set of at least $0.7n$ columns, (ii) every raw entry is a legal code of the declared format, $\lvert{\rm cd}\rvert<2^{22}$ and the row exponent is maximal, (iii) $\mathbf W^{\rm cd}=\mathrm{CD}(\mathbf W^{\rm raw})$ with the exponents of the scale row, (iv) the norm row is exact, (v) the radius rows equal $\lvert\mathbf W^{\rm cd}\rvert+\kappa m$, and (vi) the configuration rows are well formed. It is split into row-block sub-proofs (147 blocks of 7 rows for Qwen3-0.6B) that first commit their message rows; only after \emph{all} blocks are committed does one Fiat--Shamir query draw $m^{\rm s}_q=206$ column positions shared by all blocks and both trees; one column sub-proof opens these columns as hidden witnesses inside the circuit and opens both roots from \texttt{model\_\allowbreak{}id}; and the sub-proofs are linked by masked public scalars that must sum to zero. Soundness is direct: if at least $0.3n$ columns disagree with the committed messages, the shared positions miss all of them with probability at most $0.7^{206}=2^{-106.0}$; otherwise the agreeing columns are a common agreement set larger than the unique-decoding bound $(1+\vartheta)n/2$, so every committed row decodes to its committed message; a hit is caught by a nonzero bilinear form in the later batching coefficients (error $2/\lvert\mathbb E\rvert$). No correlated-agreement theorem is needed, because the messages are committed rather than extracted from a random combination. (Independent per-block column draws would certify only per-block agreement sets whose intersection could be empty; drawing one shared column set after committing all message row blocks avoids both that pitfall and any correlated-agreement assumption.) Because the shared positions depend on every block's commitment, the prover computes each block's first phase twice, regenerating masks and salts from a per-block AES-256-CTR key, so the setup proof's zero knowledge is computational. Items (i)--(vi) form the complete one-time setup relation.

\textbf{Per proof [Def].} All sampled GEMMs (both $\mathbf W^{\rm cd}$ and $\mathbf W^r$ products), the LM head and embedding lookups are reduced to \emph{hidden} tile claims at one public point in two commit rounds: a public row point $\boldsymbol\xi$, then commitments to the folded rows $\sum_{o'}\mathrm{eq}(\boldsymbol\xi,o')\mathbf W_t(o',\cdot)$ with one zero-sum per GEMM and output tile row tying them to the committed activations, then a public column point $\boldsymbol\zeta$ and the claims $c_j=\widetilde W_{t_j}(\boldsymbol\xi,\boldsymbol\zeta)$ with hidden tile indices $t_j$. Claim slots come in pairs, slot $j+J/2$ holding the radius tile of slot $j$'s tile, enforced by $t_{j+J/2}=t_j+R_{\max}/2$. A Fiat--Shamir $\gamma_{\rm hco}$ gives a hidden combination vector $\alpha_i=\sum_j\gamma_{\rm hco}^j[t_j=i]$, nonzero only on touched rows, enforced by a weighted LogUp $\sum_i\alpha_i/(\lambda-i)=\sum_j\gamma_{\rm hco}^j/(\lambda-t_j)$ (with $t_j\in[0,n_r)$, the two rational functions agree only if every $\alpha_i$ is correct, so a wrong $\boldsymbol\alpha$ passes with probability at most $(n_r+J_{\max})/\lvert\mathbb E\rvert$); the prover commits $\mathbf u=\boldsymbol\alpha^\top\mathbf W$ and proves $\sum_j\gamma_{\rm hco}^jc_j=\tilde{\mathbf u}(\boldsymbol\xi,\boldsymbol\zeta)$. Then a set $\mathcal Q$ of $m_q$ public random column positions is drawn; the static column values and Merkle paths enter the outer proof as \emph{hidden witnesses}, are checked against $\mathsf C_W$ inside the circuit, and a batched check with coefficients $\beta_{{\rm col},j}$ verifies $\sum_{j\in\mathcal Q}\beta_{{\rm col},j}\mathrm{RS}(\mathbf u)[j]=\sum_{j\in\mathcal Q}\beta_{{\rm col},j}\langle\boldsymbol\alpha,\mathbf M_{:,j}\rangle$. Positions are public, contents are not; the static codeword never appears in a public transcript. These rounds are part of each response's transcript (Section 8.3), and the LM head is handled the same way, which is how logits are recomputed in-circuit.

\textbf{Properties.} Soundness: if $\mathbf u\neq\boldsymbol\alpha^\top\mathbf W_*$, a column passes with probability at most $\vartheta+\delta_{\rm ca}$ with $\delta_{\rm ca}=0.3$ the fraction of columns outside the common agreement set certified by $\pi_{\rm setup}$, so $m_q=123$ columns give $0.55^{123}\approx2^{-106.09}$ [Paper]. Weights are extracted, straight-line under A-AROM because the column trees are checked in-circuit, and fixed at publication time, before any proof's challenges. Zero knowledge: every weight-dependent value in a transcript is a witness column covered by that proof's own fresh blinding, so the zero-knowledge bound has no opening-budget term [Paper, Spec 09 Thm 6.1]. Prover work is $O(\text{touched weights})$; MoE experts that were not activated are never read. Codeword storage is about 32 bytes per parameter, 64 with the radius matrix. Measured costs are in Section 12.7. A hidden-prompt or sampling-key rotation re-commits only $\mathsf C_{\rm cfg}$ under a fresh salt while reusing $\mathsf C_W$ and $\mathsf C_{\rm raw}$ (re-running $\pi_{\rm setup}$ in the fixed-size setup container in Stage 1; incremental setup certificates are deferred to Stage 2).

\subsection{Opening the hidden configuration inside the proof}
To prevent a prover from substituting another committed tile of the same shape (e.g., from a different layer) and to bind per-row decode exponents, score-exponent windows, layer kinds, attention windows/sinks, cost weights, the sampling key, and hidden prompts without exposing architecture metadata, we partition the non-weight configuration by update frequency [Def; Paper].

\emph{Stable tables} change only with the weights: the tile table $(\nu,e,a,b)\mapsto(t,G_t,\kappa_t)$, a vector table locating gain vectors and embedding rows, a layer table (kind, window, sinks, dimensions, RoPE, $\epsilon$, score-exponent window, contract tier), the graph, cost and port tables, and the tokenizer. They are written into $N_{\rm cr}$ configuration rows of $\mathbf W$ at profile-fixed offsets, so $\mathsf C_W$ commits them, and $\pi_{\rm setup}$ item (vi) proves once that every table has strictly increasing keys (so a lookup returns a unique value), that $G_t$ equals the scale row, that $\kappa_t$ matches the input width and contract, and that derived parameters such as the score-exponent window follow from the committed gains. In each proof, the circuit already commits the table side of every lookup as a hidden column in its first message; we add one claim slot per configuration row at the shared point, $c_r=\sum_k\sum_i\mathrm{eq}((\boldsymbol\xi,\boldsymbol\zeta),o_k+i)T_k(i)$, so the hidden-column opening ties the tables used in the circuit to the committed rows. Since the table columns are committed before $\boldsymbol\xi,\boldsymbol\zeta$ are drawn and the weights are fixed at publication, a mismatched table survives with probability at most $2\log_2T/\lvert\mathbb E\rvert$, the same round as the tile reduction; no new challenge round is needed. Each $\mathbf W^{\rm cd}$ claim slot then looks up $(t_j,\nu_j,e_j,a_j\Vert b_j,G_j,\kappa_j)$ in the tile table, with $\nu_j,e_j$ taken from the sampled block's slot (fixed before sampling) and $a_j,b_j$ public constants of the gadget's slot plan; key uniqueness makes $t_j$ unique.

\emph{Volatile configuration} (sampling key, sampling-rule parameters, hidden prompts and templates padded to a profile length $P_{\max}$) is committed as $\mathsf C_{\rm cfg}=H_{\rm in}(\texttt{dom}_{\rm cfg},\text{versions},K_{\rm prf},\mathsf{Enc}_{P_{\max}}(M^{\rm vol}))$ and opened in full in every proof by an in-circuit sponge feeding the existing in-circuit \texttt{model\_\allowbreak{}id} sponge; the key column also seeds the sampling PRF, and the prompt column is the table side of the rendering relation. Prompt or key changes therefore touch only $\mathsf C_{\rm cfg}$.

Every new value is a freshly blinded hidden column and every shape is a profile constant, so zero knowledge is unchanged. For a Qwen3-0.6B-class profile the opening costs about $3$--$4\times10^5$ data cells per proof (the configuration sponge fits in the padding of the existing Poseidon2 region), against the 0.38 G-cell placeholder in the early census; adding items (v) and (vi) over $\mathbb F_{p^3}$ brings the complete one-time setup proof $\pi_{\rm setup}$ on full \texttt{Qwen3-0.\allowbreak{}6B} to 8.67 G cells (\texttt{3,623.\allowbreak{}7 s} GPU prove, \texttt{59.\allowbreak{}9 s} CPU verify [Meas: Qwen3-0.6B, 4070TiS], Section 12.7).

\section{Certified Tolerance}
\subsection{Rounding contracts}
A contract $\mathcal C$ maps each operator kind and number format to an error-bound template from a fixed menu, with parameters [Def]. Templates bound the difference between the engine output and the operator's exact value on the \emph{actual} input:

\begin{center}\small
\begin{tabularx}{\linewidth}{>{\hsize=0.55\hsize}Y>{\hsize=1.45\hsize}Y}
\toprule
{}\textbf{Template} & {}\textbf{Statement on its domain} \\
\midrule
{}$\mathsf{Store}_k$ & {}$\hat v=v(1+\theta)+\eta$, $\lvert\theta\rvert\le u_k$, $\lvert\eta\rvert\le\eta_k$ ($u$: bf16 $2^{-8}$, fp16 $2^{-11}$, fp32 $2^{-24}$, e4m3 $2^{-4}$); $\eta_k$ covers subnormals and flush-to-zero \\
{}$\mathsf{Dot}(n;\mathcal T)$ & {}$\lvert\widehat{\langle a,b\rangle}-\langle a,b\rangle\rvert\le\gamma_h\langle\lvert a\rvert,\lvert b\rvert\rangle+\eta$, $h$ the effective depth of the actual reduction DAG \\
{}$\mathsf{Dot}^{\rm sk}$ & {}split-K: $S$ contiguous segments, partial sums stored in format $k_p$, reduced in $k_r$; radius maximized over the allowed $S$ \\
{}$\mathsf{Elem}_f$ & {}$\lvert\hat f(x)-f(x)\rvert\le u_f\lvert f(x)\rvert+\eta_f$, $\eta_f$ covering special paths (overflow-to-zero, FTZ) \\
{}$\mathsf{AttnFA}(n;B_c,S_{kv})$ & {}composite bound for online-softmax attention with key tile $B_c$ and split-KV $S_{kv}$ \\
{}$\mathsf{Reduce}(P;\mathcal T_r,k)$ & {}cross-device reduction with topology-dependent depth (ring: $P-1$) \\
{}$\mathsf{Scan}(B_c)$ & {}chunked linear attention: \textbf{[Open]}; architectures that need it are outside the completeness theorem until it is proved \\
\bottomrule
\end{tabularx}
\end{center}

Hardware or kernel vendors publish contract tables per (hardware, library version, operator, format); the provider commits its choice inside \texttt{model\_\allowbreak{}id}. A public cap $\mathcal C_{\max}(\mathrm{Prof})$ per profile bounds every committed contract, and detection power is always computed at the cap, since a provider can commit a looser contract than its hardware needs. Two cap entries came out of measurement: GEMM partial sums in fp32 (the ``R32'' tier; Section 12.3) and logits stored in fp32 (Section 7.3). Deterministic sub-computations (RoPE tables, masks, sampling noise, constrained-decoding masks) are committed by exact recipe and reproduced bit for bit, outside tolerance.

\subsection{Propagation}
For an operator $o$ with actual input $\hat x=x+e$, $e$ in the input box $\mathbf r_x$, if the contract gives $\lvert\hat o(\hat x)-o(\hat x)\rvert\le u\lvert o(\hat x)\rvert+\eta$ and an envelope $\lvert o(\hat x)-o(x)\rvert\le\Lambda_o(x,\mathbf r_x)$ holds on the whole box, then [Paper]

\[\lvert\hat o(\hat x)-o(x)\rvert\le(1+u)\,\Lambda_o(x,\mathbf r_x)+u\lvert o(x)\rvert+\eta .\]

Two consecutive identity stores give $2u+u^2$, the true error. Rows for GEMM, products of two engine values, elementwise maps, RMSNorm and LayerNorm, RoPE, MoE combination and row-split GEMM with cross-device reduction instantiate this template. For attention, a tile-by-tile decomposition separates rounding shared by numerator and denominator (score radius, exp, the \emph{same} rescaling factors) from rounding specific to each, giving, before the final division and store, a radius [Paper]

\[\mathbf r_o=\frac{\epsilon_N\sum_jp_j\lvert v_j-o\rvert+(1+\epsilon_\sigma)(\epsilon_\mu+\epsilon_\mu')\lvert o\rvert+(1+\epsilon_N)\sum_jp_j\mathbf r_{v_j}+\bar\eta\big(\sum_j(\lvert v_j\rvert+\mathbf r_{v_j})+n\lvert o\rvert\big)}{1-\epsilon_D-n\bar\eta}+n_t\eta_{\rm acc}.\]

The $\bar\eta$ term captures additive \texttt{exp} underflow, which a purely multiplicative \texttt{exp} error model would miss. A concrete example shows why it is essential: with $n=1025$ keys, $z_1=0$, $v_1=0$ and $z_j=-104.5$, $v_j=2^{60}$ for the others, IEEE fp32 rounds $e^{-104.5}\approx2^{-150.8}$ to 0, so a compliant FlashAttention kernel outputs 0 while the exact output is about $2^{-80.76}$, and the multiplicative radius is only a few $u\cdot o$. Each key's effective weight, in units of the running maximum, carries an additive error of at most $\eta_w=(n_t+1)(1+\epsilon_\mu)\eta_{\exp}+\eta_P$ (underflow of its own exponential, of up to $n_t$ rescaling factors, and of the stored probability), and $\bar\eta=e^{\bar r_z}\eta_w$ converts to units of the exact normalizer; $\eta_{\rm acc}$ is the accumulator's store floor. $\eta_{\exp}$ is declared per deployed \texttt{exp} path ($2^{-150}$ for IEEE fp32 without flush-to-zero, $2^{-126}$ for \texttt{ex2.\allowbreak{}approx.\allowbreak{}ftz}). In the counterexample the new term is at least $2^{-150}\cdot2^{70}>o$; on natural inputs it is negligible (below $2^{-104}$ for $n=32$k, $\lvert v\rvert\le10$). The other templates already carry additive floors ($\eta_k$, $\eta_{\rm dot}$, $\eta_f$). An output cap is a \emph{set constraint} checked separately, not an error radius: a cap $C$ only implies error $C+\lvert\mathrm{ref}\rvert$. Radii are sound but not minimal: shared rounding variables (e.g. $a-a$) make node-by-node propagation loose by $O(u)$.

\textbf{Why a fixed tolerance cannot work [Paper, Spec 04 Prop. 2$'$].} For $\Delta=W\tilde x$ with bf16 inputs, each output component separately can reach error $u_b(\lvert W\rvert\lvert\tilde x\rvert)_i$, so a fixed relative box fails on any component with cancellation; and for any fixed norm-form function of $(\lVert\mathrm{ref}\rVert,\lVert\Delta\rVert)$ there are inputs that make honest error exceed it. A fixed tolerance can still be empirically fine on natural activations, but adversarial prompts and batch-dependent numerics would turn refusals into a cross-tenant leak. Certified tolerance is the route to a theorem.

\subsection{Granularity: observe every GEMM operand}
A box propagated through two contractions blows up: through $\lvert W_{\rm up}\rvert$, the gate and $\lvert W_{\rm down}\rvert$, a layer-boundary radius is tens of times the true error [Paper, random model]. Measurement confirms that layer-boundary checking is vacuous (Section 12.4). The observation rule is therefore [Def]: observe the actual operand of every matrix multiply and attention (after quantization), every carried state and the residual stream; fold every other operation into the producing block as a tail. Per token and layer this is $4d+d_{ff}+2(n_q+n_{kv})d_h$ values, 6.7--7.8 times the layer-boundary set for public dense architectures [Def].

\subsection{Cost accounting: check fine, charge coarse, follow lineage}
Measurement exposed a loophole (Section 12.4): a block whose input is the prover's choice is satisfied \emph{exactly} when that input is zero ($W\cdot0=0$, attention over zero values is zero). Skipping half the layers and publishing zero operands leaves only the cheap normalization tails bad, so per-block cost accounting reports $d_C\approx10^{-4}$ for a 38\% compute saving. The fix keeps per-GEMM checks and changes accounting [Def, rule CA]: costs are charged to sublayer \emph{accounting units} (attention sublayer, MLP sublayer, each routed expert, router+combine, head+sampling); a unit's check set is every block containing one of its operators, so the normalization hosted at the tail of the previous block belongs to the next unit; attention core cost is split into per-key atoms charged to the consumer; and an atom is bad if any unit in its lineage, the units of its own column and of the read key's column down to depth $h$, is bad. This is the threshold of Section 4.4. Then $d^{(h)}_C\ge d_C$, so the threshold relation only gets stricter and honest completeness is unaffected.

\textbf{Theorem 6.1 [Paper, conditional on an empirical approximation hypothesis].} If every execution of a cheating strategy costs at least $\sum_ac_a(1-s_{\tau(a)})\mathbf 1[V_a=0]$, with $S_{\rm tol}=\max_\tau s_\tau$ the public within-tolerance saving, then a saving $s$ forces $d^{(\infty)}_C\ge(s-S_{\rm tol})/(1-S_{\rm tol})$; for finite $h$ the bound loses a factor $\Lambda_h\le2\kappa_{\rm layer}\max(1,L/(h+1))$, and this loss is tight up to constants because prefix collapse escapes any finite lineage (Spec 06 Prop. 6.6). We adopt $h=3$, which charges all measured and known local trivializations at roughly an order of magnitude in proof cost, and treat prefix collapse as an out-of-scope exotic cheat; $h=\infty$ would multiply proving cost by $\Theta(L)$ [Est]. Under rule CA the measured zero-chain skip is charged in full: $d_C=s=0.38$ [Paper, recomputed from measured per-block flags on Qwen3-0.6B].

\subsection{What certified tolerance does not guarantee}
It does not guarantee token-for-token equality of outputs across compliant executions; it does not distinguish neighbouring number formats with the same weights as a theorem (that per-GEMM checks catch fp8 is an empirical result); and it does not prove that kernels meet their contracts, which is the hypothesis H-contract (Section 10.3). Recent work shows that a poisoned checkpoint can turn benign backend rounding differences into different outputs \cite{WYL+26}; FlashProve-Next commits the deployed configuration, not an audit configuration, and checkpoint backdoors are outside verifiable inference.

\section{Sampling: Bucketed Keyed Sampling}
\subsection{The rule}
Gumbel-max sampling needs one keyed pseudo-random value per vocabulary entry and top-$p$ needs a sort; in an earlier census this vocabulary interface cost about 218 committed cells per vocabulary entry per checked step [Meas: symbolic census, i7-14700F]. The committed rule $\mathsf{Samp}^{\rm bk}_{\beta_s}$ (\texttt{fpn-samp-bk-v3}) is written in integer form, bit-identical to the reference implementation [Def]. Its public parameters are the temperature $T$, the 16-bit top-$p$ value $P_{16}=\mathrm{round}(p_{\rm top}2^{16})$ and the bucket size $\beta_s$; it commits three tables: an exponential table $E[j]=\mathrm{round}(2^{40}e^{-j/2^{12}})$ (non-increasing, with $E[J]=0$ at $J=116{,}405$), a discrete Gumbel table $\mathsf G$ with $2^{24}$ entries, and a monotone fixed-point logarithm $\mathsf{LN}$. Given fp32 logits $z$ and the allowed set $A_t\neq\varnothing$:

\begin{enumerate}
\item{} quantize, $m=\max_{v\in A_t}z_v$, $j_v=\lfloor\mathrm{fl}_{64}(\mathrm{fl}_{64}(m-z_v)\cdot c)\rfloor$ with $c=\mathrm{fl}_{64}(2^{12}/T)$, and set $w_v=E[\min(j_v,J)]$ for $v\in A_t$, else 0 (underflow is represented exactly by $E[J]=0$; these three binary64 operations are the only floating-point steps);
\item{} top-$p$ by threshold, without sorting: $S=\sum_vw_v$, $\mathrm{tgt}=\lceil SP_{16}/2^{16}\rceil$, $\theta^\star=\max\{\theta:\sum_{w_v\ge\theta}w_v\ge\mathrm{tgt}\}$, nucleus $N=\{v:w_v\ge\theta^\star\}$ (it differs from sorted top-$p$ only in taking all ties at the threshold);
\item{} partition the vocabulary into $B=\lceil V/\beta_s\rceil$ contiguous buckets with masses $W_b=\sum_{v\in b\cap N}w_v$;
\item{} choose a bucket by Gumbel-max with one keyed noise per bucket, $b^\star=\arg\max_b(\mathsf{LN}(W_b)+\mathsf G[\mathrm{idx}_b])$, ties to the smallest index, where one PRF call $\mathsf{PRF}_K(\mathsf{ctx},d_t,t,\texttt{bkt},\lfloor b/16\rfloor)$ yields 8 field elements, each split into two 24-bit table indices, covering 16 buckets;
\item{} inside the bucket, inverse CDF with one 48-bit uniform $u_{48}$ from $\mathsf{PRF}_K(\mathsf{ctx},d_t,t,\texttt{inv})$: output the first $v\in b^\star$ with prefix sum $S_v>\lfloor u_{48}W_{b^\star}/2^{48}\rfloor$.
\end{enumerate}

The nucleus has an equivalent per-token form [Lean: \texttt{BucketSampling.\allowbreak{}nucleus\_\allowbreak{}iff\_\allowbreak{}per\_\allowbreak{}token}, Spec 06 Lemma 6.7a]: with $q=2^{16}-P_{16}$, $A(v)=\sum_{u:w_u>w_v}w_u$ and $R(v)=S-A(v)$, $v\in N$ iff $w_v>0$ and $qA(v)<P_{16}R(v)$. With continuous Gumbels, a continuous uniform and real arithmetic, $\Pr[v]=w_v/\sum_{u\in N}w_u$ exactly [Paper]. With the committed tables and integer arithmetic, the rule's distribution is \emph{defined} as that of the discrete process on uniform field elements; the ideal functionality in the zero-knowledge proof passes truly random values through the same maps, so discretization adds no error term. Against float softmax truncated to the nucleus, chi-square tests ($N=40{,}000$; $\beta_s\in\{1,8,32,V\}$, top-$p\in\{1,0.9,0.95\}$, $T\in\{1,0.6\}$, with and without masks) give $p=0.43$--$0.95$ and total variation at most $3.5\times10^{-5}$ [Meas: reference implementation, i7-14700F]. Default $\beta_s=32$. $\beta_s=1$ is plain Gumbel-max; $\beta_s=V$ is a single-PRF inverse CDF, which we reject because it gives \emph{every} token an additive reachability of about the relaxation width and degrades speculative coupling. Version 1 has no top-$k$ (Qwen3's default \texttt{top\_\allowbreak{}k=20} is not applied); top-$k$ can be added as a new rule version with the same sort-free per-token form.

\subsection{Certified relaxation}
The head block supplies reference logits $z$ and radii; let $R=\max_{v\in A_t}r_{z,v}$. An admissible engine has $\lvert\hat z_v-z_v\rvert\le R$, hence $\lvert\hat m-m\rvert\le R$ and $\lvert(\hat m-\hat z_v)-(m-z_v)\rvert\le2R$, so its table index satisfies $\lvert\hat j_v-j_v\rvert\le\Delta_{\rm b}:=\lceil2Rc\rceil+2$ (two floors, one step of margin). Note the factor $2R/T$ rather than $R/T$: once logits are quantized to integer weights, the common maximum shift $\hat m-m$ no longer cancels in normalization. With $w^{\rm lo}_v=E[\min(j_v+\Delta_{\rm b},J)]$ and $w^{\rm hi}_v=E[\max(j_v-\Delta_{\rm b},0)]$, every admissible engine weight lies in $[w^{\rm lo}_v,w^{\rm hi}_v]$, \emph{including underflow}: $\hat j_v\ge J$ forces $j_v+\Delta_{\rm b}\ge J$, i.e. $w^{\rm lo}_v=0$. $\mathsf{LN}$, $\mathsf G$, the prefix sums and comparisons are committed integer operations, bit-identical in the engine and the circuit, so no additive weight floor and no logarithm ulp slack are needed; the sampler contract reduces to the three binary64 operations of step 1.

\textbf{Nucleus bounds without witnesses [Def].} Define

\[\mathrm{sure}(v):\quad w^{\rm lo}_v>0\ \wedge\ q\sum_{u\ne v:\,w^{\rm hi}_u>w^{\rm lo}_v}w^{\rm hi}_u<P_{16}\Big(w^{\rm lo}_v+\sum_{u\ne v:\,w^{\rm hi}_u\le w^{\rm lo}_v}w^{\rm lo}_u\Big),\]

\[\mathrm{poss}(v):\quad w^{\rm hi}_v>0\ \wedge\ q\sum_{u:\,w^{\rm lo}_u>w^{\rm hi}_v}w^{\rm lo}_u<P_{16}\sum_{u:\,w^{\rm lo}_u\le w^{\rm hi}_v}w^{\rm hi}_u,\]

and $N^{\rm lo}=\{\mathrm{sure}\}$, $N^{\rm hi}=\{\mathrm{poss}\}$. \textbf{Lemma 7.1} [Lean: \texttt{BucketSampling.\allowbreak{}nucleus\_\allowbreak{}sandwich}, Spec 06 Lemma 6.7$'$]: for every admissible engine, $N^{\rm lo}\subseteq\hat N\subseteq N^{\rm hi}$. The proof bounds the engine's $A(v)$ and $R(v)$ from the per-token form: any $u$ above $v$ for the engine has $w^{\rm hi}_u>w^{\rm lo}_v$, any $u$ with $w^{\rm hi}_u\le w^{\rm lo}_v$ is at or below $v$, and symmetrically for $\mathrm{poss}$. Both sets are deterministic functions of the reference logits and $R$, so no witness can fail to exist. \textbf{Lemma 7.2} [Lean: \texttt{BucketSampling.\allowbreak{}sure\_\allowbreak{}mono\_\allowbreak{}j}, \texttt{poss\_\allowbreak{}mono\_\allowbreak{}j}, Spec 06 Lemma 6.7$''$]: both predicates are monotone in $j_v$ (plateaus of $E$ included), so $N^{\rm lo}$ and $N^{\rm hi}$ are prefixes in $j$-order. In the circuit the prover points to hidden boundary tokens (the last token in each set and the first token outside it, or a public empty flag), the circuit evaluates the two predicates at those four tokens, and every vocabulary entry is classified against the boundary levels by an interval index and two range-checked differences. By monotonicity the boundary tokens exist and the checks pin the sets with zero slack [Lean: \texttt{BucketSampling.\allowbreak{}prefix\_\allowbreak{}boundary\_\allowbreak{}exists}, \texttt{boundary\_\allowbreak{}pair\_\allowbreak{}sound\_\allowbreak{}complete}].

The circuit accepts token $u$ iff $u\in A_t\cap N^{\rm hi}$, its bucket satisfies $\mathsf{LN}(W^{\rm hi}_{b(u)})+\mathsf G[\mathrm{idx}_{b(u)}]\ge\max_{b:W^{\rm lo}_b>0}(\mathsf{LN}(W^{\rm lo}_b)+\mathsf G[\mathrm{idx}_b])$ with $W^{\rm hi}_b,W^{\rm lo}_b$ the masses of $w^{\rm hi}$ on $N^{\rm hi}$ and $w^{\rm lo}$ on $N^{\rm lo}$, and $S^{\rm lo}_{u-1}2^{48}\le u_{48}W^{\rm hi}_b$ and $u_{48}W^{\rm lo}_b<S^{\rm hi}_u2^{48}$ for the in-bucket prefix sums. \textbf{Theorem 7.3} [Lean: \texttt{BucketSampling.\allowbreak{}bucket\_\allowbreak{}sampling\_\allowbreak{}complete}, Spec 06 Thm 6.8]: under H-contract and the range-certificate domain condition $2Rc<2^{52}$ (enforced in circuit by $\Delta_{\rm b}\le\texttt{DMAX}=2^{15}$), for every vocabulary size, nonempty allowed set, $T>0$, $p_{\rm top}\in(0,1]$ and logits (underflow, plateaus and ties included), the honest engine's token is always accepted. Finite tests agree: engines whose logits were perturbed within $r\in\{0,10^{-3},10^{-2},0.1\}$ ($V=4096$) and $r\in\{10^{-3},3\times10^{-2}\}$ ($V=151{,}936$) were accepted in every trial, and the torch/Triton engine matches the reference on 96 rows of real Qwen3-0.6B logits captured inside vLLM [Meas: 4070TiS].

\textbf{Why a witness-based global threshold fails.} A naive relaxation with hidden scalar thresholds $\theta^{\rm lo},\theta^{\rm hi}$ satisfying $\sum_{w^{\rm lo}\ge\theta^{\rm lo}}w^{\rm lo}\ge p_{\rm top}\sum w^{\rm hi}$ under a global factor $e^{\pm\bar\delta}$ has no witness whenever $p_{\rm top}>e^{-2\bar\delta}$ (including $p_{\rm top}=1$ for any $\bar\delta>0$), because the left-hand sum is at most $e^{-\bar\delta}\sum w$ while the right-hand target is $p_{\rm top}e^{\bar\delta}\sum w$. Even when a witness exists, $\{w^{\rm lo}\ge\theta^{\rm hi}\}$ drops the boundary token itself, shrinking that bucket's lower mass and loosening the inverse-CDF bound: on $V=4096$, top-$p=0.95$, $r=0.1$, scalar-threshold bounds admit 676 tokens per step on average whereas the witness-free per-token sandwich above admits 2.6 [Meas: reference implementation, i7-14700F].

\subsection{Manipulation band, speculative decoding and cost}
The relaxation also lets a key-holding provider choose among tokens inside the band: with $\bar\delta=\Delta_{\rm b}2^{-12}\approx2R/T+2^{-11}$, a token is reachable with probability at most about $(1+2\bar\delta)\pi_u+\bar\delta\pi_{b(u)}$ [Paper]. This saves no compute and is a declared non-goal, but the band is published with \texttt{model\_\allowbreak{}id} per temperature. With logits computed in bf16, $\bar\delta\approx0.08$--$0.24$ at $T=1$; with fp32 logits about $2.5\times10^{-3}$ at $T=1$ [Est from measured head radii], growing as $1/T$ (about $10^{-2}$ at $T=0.2$), which is why $\mathcal C_{\max}$ requires the engine to compute and sample from fp32 logits (they are never written out or committed, so this costs no online bandwidth). For wide models the row-norm (Cauchy--Schwarz) radius can be looser than $\lvert E\rvert\lvert x\rvert$ by a factor with no general bound; 70B-class radii are unmeasured. The rule is a deterministic function of prefix and noise, so coupled speculative decoding reproduces sequential selection exactly on the verification forward's logits [Lean: \texttt{Speculative.\allowbreak{}specLoop\_\allowbreak{}eq\_\allowbreak{}seqGen}, for a fixed deterministic step function]. Acceptance is lower than with per-entry Gumbel: on synthetic Gaussian logits ($V=4096$, noise scale 0.1/0.3/1.0) the per-position match rate is 0.87/0.74/0.48 at $\beta_s=32$, 0.95/0.85/0.59 at $\beta_s=1$ and 0.55/0.30/0.21 at $\beta_s=V$ [Meas: reference implementation, i7-14700F]; real-model acceptance is unmeasured. Circuit cost, counted item by item against the reference implementation, is 13--24 data cells per vocabulary entry in the asymptotic target; our Stage 1 implementation (\texttt{sampler\_\allowbreak{}zk} + \texttt{gen\_\allowbreak{}zk}) shares one \texttt{BatchedHead} weight-claim reduction across all $K=128$ padded generation slots ($k_2=105$ draws) and replaces the $2^{24}$-row monolithic Gumbel lookup with a 65k-entry piecewise-linear body plus 131k-entry exact dual-tail table, running end-to-end at full $V=151{,}936$ inside the 315-draw \texttt{Qwen3-0.\allowbreak{}6B} proof (Section 12.8). Each checked step adds 298 Poseidon2 permutations, about $2^{21.8}$ S-boxes per proof at $k_2=105$ versus $2^{27.6}$ for per-entry noise. In vLLM 0.33 on 4070TiS (Qwen3-0.6B, CUDA graphs) the rule alone costs 0.11--0.31 ms per step, 3.8--7.2\% of decode throughput at batch 1--32 [Meas].

\section{Proof System and Lifetime Knowledge Soundness}
\subsection{Two hashes and the security model}
$H_{\rm out}$ (polynomial-commitment Merkle trees and Fiat--Shamir) is evaluated only outside the circuit and is modelled as a classical programmable random oracle, instantiated with BLAKE3 or SHA-3 separate from Poseidon2 [Assume]. $H_{\rm in}$ (trace leaves and tree, \texttt{model\_\allowbreak{}id}, the weight column trees and configuration root, state anchors) is a Poseidon2 sponge evaluated \emph{inside} the zero-knowledge circuit, because opening leaves outside the circuit would reveal which addresses were sampled and which KV entries attention read. Extraction must still observe its queries, which goes beyond the plain ROM. All parts of the analysis use one model of Poseidon2: the permutation $P$ is an ideal \emph{invertible} permutation (the adversary may query $P$ and $P^{-1}$), and $H_{\rm in}$ and the sampling PRF are domain-separated sponges over it. Zero knowledge uses only the hiding of salted sponge outputs (sponge indifferentiability \cite{BDPV08}) and keyed-sponge PRF bounds; the simulator neither programs nor observes $P$. Binding uses collision and use-before-query bounds counted over permutation calls. Only extraction needs more, because the circuit checks the \emph{concrete} Poseidon2 constraints while an ideal permutation has no circuit:

\textbf{Assumption A-AROM.} (a) The in-circuit Poseidon2 gadget is an exact arithmetization of $P$; (b) every pair $(x,P(x))$ in a witness extracted from an accepted proof (trace leaves and paths, weight column trees, \texttt{model\_\allowbreak{}id}, state anchors) was obtained by an adversary query to $P^{\pm1}$ before the corresponding Fiat--Shamir query, except with probability $\Delta_{\rm arom}(Q_{\rm tot})$.

This is the situation formalized by the arithmetized random oracle model of \cite{CCGOS23}, which pairs a random oracle with an arithmetization oracle so that in-circuit statements about oracle evaluations are consistent with oracle queries; we use that model only as the framework for stating A-AROM and invoke none of its theorems (its results construct proof-carrying data). For concrete Poseidon2, A-AROM is a heuristic shared by every system that hashes inside a circuit and extracts outside. A pure-ROM alternative (R3) puts all trace values into one global zero-knowledge polynomial commitment at proving cost linear in the trace; it covers the trace side only, because opening the weight column trees outside the circuit would publish static codeword symbols (Section 5.1), so the model commitment always needs A-AROM. We do not claim QROM security.

The budget distinguishes $Q$, first queries to the \emph{challenge} domains over the lifetime (all re-salting, re-committing, abandoned and retried proofs, offline precomputation, and the verifier's first evaluations), from $Q_{\rm tot}\ge Q$, all first hash queries including honest leaf hashing. An honest provider may hash $2^{60}$ leaves in its lifetime; leaf hashing creates no challenges and cannot be used to grind, so it enters only a negligible collision term. Defaults: $Q=2^{64}$, $Q_{\rm tot}\le2^{80}$, $\varepsilon=2^{-40}$.

\subsection{Straight-line extractor}
Plain soundness is vacuous here: almost every root ``has'' a preimage. We need knowledge soundness with partial-witness semantics. The extractor $\mathcal E_{\rm SL}$ reads the lifetime table of hash queries and answers, without programming or rewinding, and for an accepted $(X,\pi)$ walks from \texttt{model\_\allowbreak{}id} (once per model, at publication: configuration root, weight column trees, unique decoding $W_*$ guaranteed by the accepted $\pi_{\rm setup}$), $R$, $\mathsf{cm}_{\rm glob}$ and the anchors to recover the configuration, every object or $\bot$, the global columns (unique decoding within the code's radius) and the anchor openings [Def]. It makes zero extra oracle queries and costs $O(Q_{\rm tot}\log Q_{\rm tot})$ to index. Counting permutation calls rather than treating multi-block sponge hashes as single oracle queries, its failure events are (B1) two calls whose answers agree on the digest part, (B2) an answer equal to a digest that already appeared in an earlier call input without being an earlier answer, and (B3) a sponge inner collision, an answer of $P$ or $P^{-1}$ whose capacity part hits an earlier chaining value; digests and capacity are four Goldilocks elements (range $p^4\approx2^{256}(1-2^{-30})$), and with at most $\alpha<2^9$ digest fields per call input,

\[\Delta_{\rm SL}\le(\alpha+2)Q_{\rm tot}^2/p^4<1.01\cdot 2^{-87}\le 2^{-86}\qquad[\text{Paper}]\]

[Lean: \texttt{ProtocolBridge.\allowbreak{}delta\_\allowbreak{}SL\_\allowbreak{}p4\_\allowbreak{}default}]. Without B1/B2, the witness extracted with the full table equals the one extracted at the moment of the response's sampling query, so violation fractions are fixed before sampling (\emph{prefix preservation} is a property, not an interface assumption).

\subsection{Transcript and stratified sampling}
Each response has its own Fiat--Shamir transcript, with no cross-response (batch) challenge [Def]: (1) $m_0=(X_0,R,\mathsf{cm}_{\rm glob})$ where $\mathsf{cm}_{\rm glob}$ commits the \texttt{model\_\allowbreak{}id} opening, tokens, slot table, automaton columns, generation-step table and state openings; (2) sampling challenge; (3) commitments to the local witnesses of every block in the sampled atoms' lineages; (4) the hidden-column-opening rounds of Section 5: the row point $\boldsymbol\xi$, the folded rows, the column point $\boldsymbol\zeta$ and the committed claim column (tile claims and configuration-row claims), $\gamma_{\rm hco}$ followed by commitments to $\boldsymbol\alpha$ and $\mathbf u$, the column positions $\mathcal Q$ followed by the hidden columns and paths, then $\beta_{\rm col}$ and the column-check sumcheck; (5) LogUp challenges for all lookups (including the tile table); (6) zero-check points, Libra mask combination $\mu_{\rm L}$, sumcheck rounds, batching; (7) Basefold folding challenges $\chi_j$ and query positions. The setup proof $\pi_{\rm setup}$ is a separate per-model transcript (every row block's message rows committed, then one shared set of column positions drawn from all block roots, the hidden-column sub-proof, the batching and link coefficients, then the algebraic rounds of every sub-proof and the zero-sum check of the link scalars) whose first queries count against the same $Q$. Each challenge hashes the previous challenge and the new message, and the sampling challenge hashes all of $X_0$ including the full visible response, so visible tokens cannot be chosen after sampling.

Sampling expands the challenge into $\sum_jk_{1,j}+k_2$ uniform $n_{\rm smp}=192$-bit integers and maps each into its pool by $J=\lfloor UC_j/2^{192}\rfloor$, proved with a multi-limb division; the sampled cost unit selects a unique slot and cost atom through the committed prefix sums, and the atom's lineage is opened [Def]. The circuit always expands exactly $\sum_jk_{1,j}$ block checks and $k_2$ step checks, never deduplicates, never switches to full checking for small pools, and handles empty strata with a private flag, so the number of checks reveals nothing. The finite map adds bias $a_{\rm map}\le(\sum_jk_{1,j}C_{\max}+k_2N_{\max})/2^{192}\le2^{-107}$ [Lean: \texttt{bucketCount\_\allowbreak{}bounds}, \texttt{multi\_\allowbreak{}pool\_\allowbreak{}map\_\allowbreak{}le}, \texttt{amap\_\allowbreak{}default}].

\subsection{Round-by-round analysis}
A naive argument bounds the bad-accept probability for each fixed sampling prefix and then takes a union over $Q$ prefixes. This is wrong for multi-round Fiat--Shamir: fixing one bad prefix costs one sampling query, after which the adversary can spend the rest of its budget grinding later sumcheck or Basefold challenges \cite{BCS16,CCH+19,CMS19}. A single-round union bound applies only to one-challenge protocols and cannot be used across multi-round Fiat--Shamir reductions.

We instead define, at each challenge query, a \emph{state} consisting of round label, all previous messages and all committed objects extracted from the table at that moment, fixed before the answer. A state is \emph{doomed} if the remaining claim is false: at sampling, $\mathsf{BadAccept}$ holds for the extracted witness; after sampling, the conjunction of $E$, $\mathrm{Chain}$, $V_a=0$ for sampled atoms, $S_t$ for sampled steps and proximity of committed oracles is false; in the hidden-column-opening rounds, the claim that every tile claim equals the extracted weights' evaluation (then its $\gamma_{\rm hco}$-combination, then the queried columns' consistency) is false; later, the reduced claim of the compiled protocol is false. Four paper premises connect the protocol to this abstraction: \textbf{P-final} (the verifier rejects every doomed complete transcript), \textbf{P-compile} (the constraint system is compiled correctly and each reduction keeps false claims false except for lucky challenges), \textbf{P-loc} (sampled atoms' indicators are decided by exactly the objects of their lineage opened in the circuit), and \textbf{P-rbr} (per-round escape probabilities):

\begin{center}\small
\begin{tabularx}{\linewidth}{YYY}
\toprule
{}\textbf{Round} & {}\textbf{Escape probability from a doomed state} & {}\textbf{Evidence} \\
\midrule
{}sampling & {}$\max\{\max_j(1-\sigma)^{k_{1,j}},(1-\rho)^{k_2}\}+a_{\rm map}$ & {}[Lean: \texttt{multi\_\allowbreak{}pool\_\allowbreak{}map\_\allowbreak{}le}, \texttt{two\_\allowbreak{}pool\_\allowbreak{}all\_\allowbreak{}good\_\allowbreak{}le}] \\
{}zero-check point & {}$n_{\rm var}d/\lvert\mathbb E\rvert$ & {}[Paper] Schwartz--Zippel \\
{}sumcheck round $r$ & {}$d_r/\lvert\mathbb E\rvert$ & {}[Lean: \texttt{ProtocolBridge.\allowbreak{}sumcheck\_\allowbreak{}round\_\allowbreak{}esc\_\allowbreak{}le}, \texttt{Sumcheck.\allowbreak{}agree\_\allowbreak{}card\_\allowbreak{}le}] \\
{}LogUp ($w$ columns, $n_v$ queries, $n_t$ rows) & {}$(n_t(w-1)+n_v+n_t)/\lvert\mathbb E\rvert$ & {}[Lean: \texttt{logup\_\allowbreak{}multicol\_\allowbreak{}card\_\allowbreak{}le}] \\
{}Libra combination / batching ($N_{\rm cl}$ claims) & {}$1/\lvert\mathbb E\rvert$ / $(N_{\rm cl}-1)/\lvert\mathbb E\rvert$ & {}[Paper] \\
{}Basefold fold $j$ / query ($s$ positions) & {}$n_j/\lvert\mathbb E\rvert$ / $(1-\delta_{\rm bf})^s$ & {}[Assume A-PCS; arithmetic Lean: \texttt{basefold\_\allowbreak{}qry\_\allowbreak{}esc\_\allowbreak{}le}] \\
{}HCO two-round tile reduction ($\boldsymbol\xi$, $\boldsymbol\zeta$; includes configuration-row claims) / $\gamma_{\rm hco}$ / $\beta_{\rm col}$ & {}$2\log_2T/\lvert\mathbb E\rvert$ / $J_{\max}/\lvert\mathbb E\rvert$ / $m_q/\lvert\mathbb E\rvert$ & {}[Paper] \\
{}HCO weighted LogUp for $\boldsymbol\alpha$ & {}$(n_r+J_{\max})/\lvert\mathbb E\rvert$ & {}[Paper] \\
{}HCO column query ($m_q=123$) & {}$(\vartheta+\delta_{\rm ca})^{m_q}=0.55^{123}\approx2^{-106.09}$ & {}[Paper; arithmetic Lean: \texttt{hco\_\allowbreak{}col\_\allowbreak{}esc\_\allowbreak{}le}] \\
{}\texttt{TC} history-matrix column query ($m_q^H=268$) & {}$(1-\delta_H)^{m_q^H}=0.76^{268}\approx2^{-106.1}$ & {}[Paper; arithmetic Lean: \texttt{tc\_\allowbreak{}col\_\allowbreak{}esc\_\allowbreak{}le}] \\
{}Constrained-decoding mask MLE check ($\mathbf r_v\in\mathbb E^{v_b}$) & {}$v_b/\lvert\mathbb E\rvert\le 2^{-186}$ & {}[Paper; Lean: \texttt{ConstrainedMask}] \\
{}$\pi_{\rm setup}$ joint column query ($m^{\rm s}_q=206$, after all blocks are committed) & {}$0.7^{206}\approx2^{-106.0}$ & {}[Paper; arithmetic Lean: \texttt{setup\_\allowbreak{}col\_\allowbreak{}esc\_\allowbreak{}le}] \\
{}$\pi_{\rm setup}$ links: column bilinear form / exponent and norm links & {}$2/\lvert\mathbb E\rvert$ / $20/\lvert\mathbb E\rvert$ & {}[Paper] \\
\bottomrule
\end{tabularx}
\end{center}

In multi-column LogUp, the polynomial-fingerprint collision term is $n_t(w-1)$ rather than $w-1$, because an invalid query tuple can collide with any of the $n_t$ distinct table rows; an explicit construction with roots of unity attains exactly $n_t(w-1)$ bad $\gamma$ [Paper; tightness checked exhaustively for small primes, Meas: Mac M1, Python]. Parallel sub-claims in one round (e.g. the $k$ sampled blocks' sumchecks sharing challenges) do not multiply the escape probability. A-PCS carries two explicit conditions: the foldable code is Reed--Solomon, so the unique-decoding radius uses the MDS distance (a random foldable code \cite{ZCF24} would need its own, smaller, distance bound), and $\delta_{\rm bf}$ is set against the \emph{effective} rate after zero-knowledge blinding; for rate $1/4$ plus $1/32$ blinding this raises the query count from about 157 to about 168.

\subsection{Main theorem and parameters}
\textbf{Theorem 8.1 (round-by-round lifetime knowledge soundness).} Under A-ROM, A-AROM (R3 may replace it on the trace side only), A-PCS, and P-final, P-compile, P-loc, P-rbr, for every lifetime adversary

\[\begin{aligned}\Pr[\mathsf{BadAccept}]&\le\Delta+\mathbb E\Big[\sum_q\varepsilon_{\tau(q)}\Big]\le\Delta+Q\,\bar\varepsilon_{\rm rbr},\\ \bar\varepsilon_{\rm rbr}&=\max\Big\{\max\{\max_j(1-\sigma)^{k_{1,j}},(1-\rho)^{k_2}\}+a_{\rm map},\ \bar\varepsilon_{\rm alg}\Big\},\end{aligned}\]

with $\Delta=\Delta_{\rm SL}+\Delta_{\rm arom}$ and $\bar\varepsilon_{\rm alg}$ the maximum non-sampling escape probability. [Paper; counting skeleton Lean: \texttt{RoundByRound.\allowbreak{}rbr\_\allowbreak{}lifetime\_\allowbreak{}soundness}, expectation form \texttt{rbr\_\allowbreak{}lifetime\_\allowbreak{}soundness\_\allowbreak{}sum}.]

\emph{Proof sketch.} Fix the adversary's coins and lazily sample both oracles; charge B1, B2 and the AROM loss to $\Delta$. Otherwise, by prefix preservation an accepted bad statement is doomed at its sampling query, and by P-final it is not doomed at the end, so some first challenge query moves it from doomed to not doomed. That state is fixed before the uniform answer, so the escape probability is at most $\varepsilon_\tau$; summing over at most $Q$ first queries gives the bound. Adaptivity is harmless because states may depend on all earlier answers. The sum form holds in expectation, since the round of query $j$ may itself be adaptive; this subtlety was found during formalization. $\blacksquare$

In Lean (\texttt{ProtocolBridge.\allowbreak{}lean}), the bridge \texttt{hpath} (every accepted bad statement hits the failure event or escapes a doomed state at some first query) is machine-checked from an abstract protocol specification \texttt{ProtocolRBRSpec} encapsulating prefix-preserving extraction (\texttt{P-ext}), lineage-local compiled constraint satisfaction (\texttt{P-compile} + \texttt{P-loc}, \texttt{compiled\_\allowbreak{}cs\_\allowbreak{}doomed\_\allowbreak{}of\_\allowbreak{}bad\_\allowbreak{}sample}), and deterministic final verification (\texttt{P-final}), yielding the composed theorem \texttt{fpn\_\allowbreak{}protocol\_\allowbreak{}lifetime\_\allowbreak{}soundness} ($\Pr[\mathsf{BadAccept}]\le 2^{-40}$ at $Q\le 2^{64}$); concrete circuit-builder instantiation of \texttt{CompiledCS} and Basefold proximity soundness (\texttt{A-PCS}) enter as explicit hypotheses.

\textbf{Sample counts.} Fix public auxiliary bounds with $\bar\Delta+Q\bar a<\varepsilon$ and set $\varepsilon_{\rm sp}=\varepsilon-\varepsilon_{\rm aux}$. Then

\[k(s,Q,\varepsilon_{\rm sp})=\Big\lceil\frac{\ln(Q/\varepsilon_{\rm sp})}{-\ln(1-s)}\Big\rceil,\qquad k_{1,j}=k(\sigma,\cdot),\ k_2=k(\rho,\cdot)\]

is exactly the least integer with $Q(1-s)^k\le\varepsilon_{\rm sp}$ [Lean: \texttt{kSuff\_\allowbreak{}iff\_\allowbreak{}kReq\_\allowbreak{}le}]. We allocate $\varepsilon_{\rm sp}=\varepsilon/2$, $\bar\Delta\le\varepsilon/4$, $\bar a=\max(a_{\rm map},\bar\varepsilon_{\rm alg})\le\varepsilon/(4Q)$; the total is then at most $\varepsilon$ [Lean: \texttt{total\_\allowbreak{}le\_\allowbreak{}rbr}, \texttt{default\_\allowbreak{}budget\_\allowbreak{}rbr}, \texttt{rbr\_\allowbreak{}default\_\allowbreak{}lifetime}].

\begin{center}\small
\begin{tabularx}{\linewidth}{YYYYYY}
\toprule
{}\textbf{$\sigma$} & {}\textbf{$\rho$} & {}\textbf{$Q=2^{30},\varepsilon=2^{-20}$} & {}\textbf{$Q=2^{30},\varepsilon=2^{-40}$} & {}\textbf{$Q=2^{64},\varepsilon=2^{-20}$} & {}\textbf{$Q=2^{64},\varepsilon=2^{-40}$} \\
\midrule
{}0.5 & {}0.5 & {}(51, 51) & {}(71, 71) & {}(85, 85) & {}(105, 105) \\
{}0.5 & {}0.2 & {}(51, 159) & {}(71, 221) & {}(85, 265) & {}(105, 327) \\
{}0.5 & {}0.1 & {}(51, 336) & {}(71, 468) & {}(85, 560) & {}(105, 691) \\
{}0.3 & {}0.3 & {}(100, 100) & {}(138, 138) & {}(166, 166) & {}(205, 205) \\
{}0.2 & {}0.2 & {}(159, 159) & {}(221, 221) & {}(265, 265) & {}(327, 327) \\
{}0.1 & {}0.1 & {}(336, 336) & {}(468, 468) & {}(560, 560) & {}(691, 691) \\
\bottomrule
\end{tabularx}
\end{center}

\emph{Table: $(k_{1,j},k_2)$, samples per computation stratum and for the generation pool, with replacement. All 16 cells of the full table (6 shown) are checked exactly in Lean, including that $k-1$ does not suffice [Lean: \texttt{table\_\allowbreak{}Q64\_\allowbreak{}e40}, \texttt{k105\_\allowbreak{}sufficient}, \texttt{k104\_\allowbreak{}insufficient}, and siblings].}

At the default point one needs 105 samples per pool (104 would be the ideal if the whole budget went to the miss event, and 104 per pool provably fails under our allocation [Lean: \texttt{k104\_\allowbreak{}fails\_\allowbreak{}rbr}]); with $\Pi_2$ that is $3\times105=315$ samples [Lean: \texttt{three\_\allowbreak{}pool\_\allowbreak{}default}].

\subsection{Field choice}
The budget requires $\bar\varepsilon_{\rm alg}\le2^{-106}$, i.e. every per-round numerator $D_\tau\le2^{-106}\lvert\mathbb E\rvert$. Over Goldilocks $p=2^{64}-2^{32}+1$, $\mathbb F_{p^2}$ allows only $D_\tau\le2^{22}-1$ [Lean: \texttt{field\_\allowbreak{}p2\_\allowbreak{}iff}], while the slot-table lookup alone has a fingerprint numerator around $2^{31.8}$ and committed codewords have length around $2^{32}$ [Lean: \texttt{slot\_\allowbreak{}table\_\allowbreak{}numerator}, \texttt{codeword\_\allowbreak{}p2\_\allowbreak{}insufficient}]. All challenges, including the hidden-column-opening, trace-commitment, constrained-mask, and setup rounds, are therefore drawn from the cubic extension $\mathbb F_{p^3}=\mathbb F_p[u]/(u^3-2)$, which allows $D_\tau\le2^{85}$ [Lean: \texttt{field\_\allowbreak{}p3\_\allowbreak{}bits}]; under the profile bound on every lookup ($n_v,n_t\le2^{48}$, $w\le8$, at most 16 lookups) the LogUp numerator stays below $2^{52}$; Basefold folding in the unique-decoding regime then contributes at most $2^{-149}$ per round [Lean: \texttt{fold\_\allowbreak{}p3\_\allowbreak{}bits}]. ``128-bit extension fields suffice'' is false for this protocol unless every round is re-budgeted with in-round repetition. Both our one-time setup prover and our end-to-end response prover execute natively over $\mathbb F_{p^3}$ with Goldilocks width-12 Poseidon2 constants matching the Plonky3 reference (Sections 12.7--12.9).

\section{Zero Knowledge with Explicit Leakage}
\subsection{Leakage function}
$\mathcal L=\mathcal L_{\rm API}\cup\mathcal L_{\rm pub}$ [Def]. $\mathcal L_{\rm API}$ is what OpenAI- and Anthropic-style APIs already return: the request (including client nonce), visible content, finish reason, usage counts including reasoning-token counts (hence hidden chain-of-thought \emph{length}, which the APIs already reveal), cache-hit counts, service tier, backend fingerprint, stream timing produced by an engine \emph{without} FlashProve-Next hooks, and API-level errors. $\mathcal L_{\rm pub}$ is: model-update events and the epoch each request binds to, the public profile (field and commitment parameters, $\mathcal C_{\max}$, strata and sample counts, lineage depth, size upper bounds, fixed proof-container size), the public service policy (refusal predicate and codes, state machine), and a public outage log of declared outage windows that apply uniformly to all tenants in a declared partition. Deliberately excluded: any global request counter or client-visible monotone sequence number, batch composition, worker assignment. The last two items in $\mathcal L_{\rm pub}$ (public service policy, public outage log) are explicitly permitted as secret-independent protocol objects (Section 2.5), consistent with the requirement that resource failures be distinguishable from numerical refusals. The number of generated steps $N$ is in $\mathcal L_{\rm API}$ in the standard mode, because usage counts including reasoning tokens determine it; it is protected only in the prefix mode, where the API does not return full usage.

\subsection{Envelope objects}
Everything the proof system adds is an envelope object that the simulator generates [Def]: $N_s$ and \texttt{proof\_\allowbreak{}id} (fresh uniform), $R$ and $\mathsf{St}_{\rm new}$ (sponge outputs on a fresh root salt and fresh anchor randomness respectively, indistinguishable from uniform when the permutation is ideal, even when every trace leaf is padding), the fixed-length AEAD blob, \texttt{model\_\allowbreak{}id} and a fixed-size setup certificate (weight updates and configuration-only updates use the same container), $\mathsf{cm}_{\rm glob}$ and the proof bytes (fixed size), and release times, failure codes and endpoint responses (deterministic functions of $\mathcal L$, the outage log and the client's own actions). Stratum sizes, empty-stratum flags, sampled positions and lineages live inside the circuit and never appear in the transcript.

\subsection{Simulator and theorem}
The simulator first draws all Fiat--Shamir challenges, then runs the IOP-level honest-verifier simulator to sample a \emph{joint} view satisfying every verifier equation (sumcheck consistency, final evaluations, fold consistency, repeated queries), then builds each committed oracle's salted Merkle tree honestly bottom-up with the simulated values at queried leaves and uniform strings elsewhere, so openings match roots without programming any Merkle node; it programs $H_{\rm out}$ only at the transcript prefixes, aborting on collisions [Paper].

\textbf{Theorem 9.1 (adaptive multi-proof multi-tenant ZK, transcript scope).} Assume $H_{\rm out}$ is a programmable random oracle (BLAKE3/SHA-3), the Poseidon2 permutation is an ideal invertible permutation with $H_{\rm in}$ and the sampling PRF domain-separated sponges over it (A-AROM is not used), the IOP masks are full rank with budget to spare, the service front end follows rules R1--R13 of Section 10.4, the setup certificate is zero-knowledge and the blob AEAD is IND\$-CPA. Then

\[\begin{aligned}\delta\le{}&N_\pi\varepsilon_{\rm deg}+\frac{q_HN_{\rm prog}+N_{\rm prog}^2}{2^{256}}+\frac{q_PN_{\rm com}}{2^{255}}+N_\pi J\varepsilon_{\rm pole}+\frac{N_{\rm req}^2}{2^{257}}\\&+\mathsf{Adv}^{\rm PRF}+N_{\rm mid}\delta_{\rm setup}+\mathsf{Adv}^{\rm AEAD}+\delta_{\rm num}+\delta_{\rm flt},\end{aligned}\]

about $2^{-88}+N_\pi G_{\rm grp}\mathsf{Adv}^{\rm PRG}+N_{\rm mid}\delta_{\rm setup}+\mathsf{Adv}^{\rm AEAD}+\delta_{\rm num}+\delta_{\rm flt}$ at lifetime parameters (when streaming mask regeneration is enabled, each commitment group derives its blinding rows from an ephemeral CSPRNG-seeded AES-256-CTR key erased upon proof release, contributing $G_{\rm grp}\mathsf{Adv}^{\rm PRG}$, while $\delta_{\rm setup}$ includes $147\cdot\mathsf{Adv}^{\rm PRG}$) ($N_\pi\le2^{48}$ proofs, $q_H,q_P\le2^{80}$), dominated among statistical terms by the retry-selection term $N_\pi J\varepsilon_{\rm pole}$ (the PRF term is about $2^{-95}$). [Paper; single-proof masking core Lean: \texttt{Mask.\allowbreak{}masked\_\allowbreak{}view\_\allowbreak{}independent}, \texttt{MaskJoint.\allowbreak{}joint\_\allowbreak{}tensor\_\allowbreak{}surjective}, \texttt{libra\_\allowbreak{}transcript\_\allowbreak{}independent}, \texttt{fold\_\allowbreak{}mask\_\allowbreak{}independent}, \texttt{multiround\_\allowbreak{}independent}.]

The hybrid order matters: proofs are replaced by simulated proofs first (which use only the public statement), then hiding commitments by uniform strings, then failure events by their deterministic public function, and only last the PRF by a random function. At that point the key appears only in the engine's sampling, so the PRF reduction needs neither the key nor a commitment to it. Composition is by induction over the release order: each proof's mask is fresh and independent of all prior public history, and mask offsets are \emph{prefix-measurable} (depend strictly on earlier messages $j<i$ of the same proof), under which each mask tuple induces a unique transcript and uniform masks push forward to the exact uniform distribution [Lean: \texttt{MaskJoint.\allowbreak{}causal\_\allowbreak{}transcript\_\allowbreak{}existsUnique}, \texttt{multiround\_\allowbreak{}independent\_\allowbreak{}causal}]. The theorem covers the transcript and release timing; stream timing additionally needs that the hooks' latency is a function of public data (assumption H-hook).

\subsection{The ``can it prove'' bit}
If an honest execution cannot be proved on some requests because of a numerical out-of-tolerance event, that event depends on hidden activations and no simulator can reproduce it. With a per-response numerical refusal rate of $2^{-20}$, an adversary sending $2^{20}$ requests gets constant advantage [Paper, Spec 03 Cor. 4.3]. Strict zero knowledge therefore requires numerical refusal to be \emph{zero} outside $\mathcal L$, which certified tolerance delivers under H-contract (Section 10.3). Any fallback where failures depend only on public data is equivalent to zero numerical failure (Spec 04 Prop. 5). In Theorem 9.1, $\delta_{\rm num}$ is the probability that some adversary-visible request fails numerically; it is 0 under H-contract and cannot be removed by any outage policy.

\subsection{Keys and secrets}
$K_{\rm prf}$ (256 bits) is fresh per \texttt{model\_\allowbreak{}id}, committed in the configuration so the provider cannot switch keys per request, and rotated before $2^{72}$ permutation calls; leaf salts are per request; masks are per attempt, erased after release, and must never be restored from snapshots (mask reuse lets two proofs be subtracted). Any secret leaking leaves soundness untouched, because the soundness adversary already holds every secret; leaking $K_{\rm prf}$ turns each known response into constraints on logits and makes future noise predictable, leaking salts turns $R$ or \texttt{model\_\allowbreak{}id} into an offline-checkable hash of guesses [Paper].

\section{Completeness}
\subsection{Three parts}
The failure probability per response is $\varepsilon_c\le\varepsilon_{\rm alg}+\varepsilon_{\rm num}+\varepsilon_{\rm del}$ [Def]: algebraic completeness (given a compliant execution and enough resources, the prover produces an accepted proof), numerical completeness (the real execution lies in the tolerance set: $E_{\rm ctr}\subseteq E_{\rm Tol}$), and on-time delivery. The request domain is everything the API accepts within the capability matrix, with no content restriction; the padded length bound covers hidden injection plus maximum context plus maximum output, so ``too long'' refusals do not occur.

\subsection{Algebraic completeness}
\textbf{Theorem 10.1 [Paper].} For any request and compliant execution satisfying the range certificate, $\varepsilon_{\rm alg}\le\varepsilon_{\rm pole}^J$ over $J$ attempts. The only imperfect step is a LogUp pole ($\lambda=-f_\gamma(\cdot)$) [Lean: \texttt{LogUp.\allowbreak{}pole\_\allowbreak{}card\_\allowbreak{}le}; completeness off poles \texttt{LogUp.\allowbreak{}logup\_\allowbreak{}complete}]; a retry re-blinds $\mathsf{cm}_{\rm glob}$, changing $m_0$ and hence every subsequent challenge. Reference traces exist and are unique [Lean: \texttt{StdAlg.\allowbreak{}trace\_\allowbreak{}exists}, \texttt{trace\_\allowbreak{}unique}, \texttt{StdAlgOps.\allowbreak{}topK\_\allowbreak{}exists}, \texttt{ArchGeneral.\allowbreak{}sbfp\_\allowbreak{}shift\_\allowbreak{}unique}, \texttt{BoundedLift.\allowbreak{}signedLift\_\allowbreak{}unique\_\allowbreak{}of\_\allowbreak{}bounded}]. Under the profile bound on lookups (each $n_v,n_t\le2^{48}$, at most 16 lookups, total at most $2^{53}$ entries), $\varepsilon_{\rm pole}<2^{-138}$. Honest proving consumes at most $N_\pi J r_{\rm FS}\le2^{60}$ challenge queries over the lifetime, inside $Q=2^{64}$; honest queries are not exempt.

\textbf{Range certificate [Paper, Spec 04 Lemma 1.1$'$, Spec 05 Prop. 2.1].} At publication the provider computes from the weights a private table bounding every object port and every \emph{internal} intermediate (partial sums, sums of squares, exponent arguments, gate products, online-softmax accumulators, recurrent state) in the data type that actually stores it. The induction is non-circular: order-independent $\ell_1$ bounds on partial sums imply no overflow step by step, which places inputs in the contract's domain, which then gives the output error bound. Recurrent state grows linearly with sequence length (when incoming keys $k$ are orthogonal to existing state rows, $\lVert S'\rVert^2=\lVert S\rVert^2+\lVert v\rVert^2$). Configurations whose certificate exceeds the format are rejected once, publicly, at publication, not per request.

\subsection{Numerical completeness and H-contract}
\textbf{Assumption H-contract.} The deployed kernels satisfy the committed per-operator contract $\mathcal C\preceq\mathcal C_{\max}$ on their domains, for this hardware, kernel library and configuration.

\textbf{Theorem 10.2 (certified completeness) [Paper].} Under H-contract and the range certificate, every execution in $E_{\rm ctr}$ satisfies $\mathrm{Tol}_b$ on every block whose operators have proven propagation rules, and every sampling step satisfies the relaxed sampling relation (Theorem 7.3). \textbf{Corollary 10.3.} For all requests in the domain, \emph{adversarial prompts included}, $\varepsilon_{\rm num}=0$; batch composition, kernel shape, tensor-parallel degree, reduction order, speculative decoding and prefill/decode disaggregation are all just members of $E_{\rm ctr}$, so numerical refusals cannot leak across tenants. Architectures needing the $\mathsf{Scan}$ template are excluded until it is proved. Both the additive \texttt{exp} underflow term in Section 6.2 and the witness-free per-token nucleus sandwich in Section 7.2 are essential for this universal statement; scoring all 392,928 honest blocks of Qwen3-0.6B and Llama-3.2-1B-Instruct under the complete contract confirms zero violations (Section 12.1).

H-contract only affects completeness: if it fails, honest proofs fail; cheating does not pass. It is a per-operator statement, testable by exact-arithmetic differential testing on adversarially constructed operator inputs, and our measurements both support it in one configuration and falsify it in another (Section 12.2--12.3). Note that a provider whose kernels violate \texttt{H-contract} cannot preserve strict zero knowledge merely by declaring the resulting failure as a public outage: an outage \emph{triggered by} a failing proof on a live request is a function of that request's hidden computation, so publishing it in the outage log only moves the leaked bit to another channel, and the ideal functionality, which runs no tolerance checks, cannot simulate it. Outages are secret-independent only when the decision depends on public data alone (for example, a kernel bug found by offline differential testing, or a scheduled withdrawal). A numerical failure on a live request is a $\delta_{\rm num}$ event: the proof fails, the incident is disclosed, and the configuration is withdrawn or fixed through a public \texttt{model\_\allowbreak{}id} update, which bounds the leak to requests before the update. Under H-contract violations we therefore do not claim strict zero knowledge.

\subsection{Delivery and failure state machine}
On-time delivery is not a theorem. Under an engineering assumption H-resource (proof slots reserved at admission, a fixed-shape circuit with worst-case time $\Delta_{\rm rel}$, trace persisted on several replicas before the stream-end event, independent retries on stateless workers with failure probability $\phi$), $\varepsilon_{\rm del}\le\phi^J$, e.g. $2^{-20}$ for $\phi=2^{-5}$, $J=4$ [Paper; $\phi$ and $\Delta_{\rm rel}$ Open]. A state machine (admit, stream, abort, commit, cut, prove, ready, released, unavailable, expired) fixes observable behaviour [Def]: $N_s$ and \texttt{proof\_\allowbreak{}id} are sent in the first stream event; the proof endpoint answers only \texttt{pending}, \texttt{proof}, \texttt{unavailable}, \texttt{policy(code)} or \texttt{none}, with never-issued, aborted and expired all answering \texttt{none}; proofs are released at fixed deadlines even if ready early; and rule R9, \emph{declarative fail-closed}, handles every unrecoverable \emph{resource} fault (crash, hardware, storage, timeout; secret-independent by nature) by publishing an outage window that covers it, never by failing an individual request. Numerical failures are not routed to R9 (they are handled as in Section 10.3), so a single adversarial prompt that triggers a kernel contract bug cannot turn into a partition-wide outage. Under R1--R13 every external event is a deterministic function of $(\mathcal L_i,\mathcal O,\text{client actions},\text{fresh randomness})$, including correlated failures, capturing correlated infrastructure faults cleanly [Paper, Spec 04 Thm 4, Lemma 6]. After a cancellation the endpoint returns no content; the client declares the prefix it received (count and hash) and the provider proves only that prefix.

\section{Production Mechanisms and API Semantics}
\subsection{Canonical logical graph and implementation templates}
Physical execution differs from the logical graph: tensor-parallel re-reductions, online rescaling in attention, chunking, fusion, MLA absorption all change the operator DAG inside a block, and exact-ring identities say nothing about floating-point error. FlashProve-Next therefore separates two layers [Def]. \textbf{Logically}, a request's trace is the unique unrolling of the committed canonical graph on the request's data; addresses, successor function, costs and strata do not depend on batch, parallel degree, placement, cache hits, prefill/decode split, chunk sizes or kernel choice, so every schedule yields the same blocks, objects, costs and steps (Theorem 3.1 of the computation model, by definition) [Paper]. \textbf{Numerically}, each node declares a set of \emph{implementation templates}: a family of allowed in-block DAGs, a parameter set, an error template from the contract menu, a domain, and a closure lemma stating that any DAG in the family within the domain stays within the template radius (maximized over parameters). The allowed execution set is all physical executions whose every block is covered by a declared template with kernels meeting H-contract. \textbf{Theorem 11.1 [Paper, conditional]}: then every block satisfies $\mathrm{Tol}_b$ and every step $S_t$. An implementation not covered by a template only causes proving failure and must be declared before deployment or disabled; contracts cannot be loosened after the fact.

\begin{center}\small
\begin{tabularx}{\linewidth}{YYY}
\toprule
{}\textbf{Mechanism} & {}\textbf{Covering template} & {}\textbf{Status} \\
\midrule
{}continuous batching, kernel autotuning & {}$\mathsf{Dot}$/$\mathsf{Dot}^{\rm sk}$ maximized over allowed splits; row independence & {}[Paper] \\
{}chunked prefill (attention), split-KV & {}$\mathsf{AttnFA}$: each key sees at most $\lceil n/B_c\rceil+S_{kv}$ rescalings, any order; additive underflow per rescaling & {}[Paper]; [Meas: 0/392,928 bad blocks] \\
{}prefix caching, PagedAttention & {}objects reuse original leaves; cache keyed by \texttt{model\_\allowbreak{}id} & {}no new obligation \\
{}TP column / row split + all-reduce & {}$\mathsf{Dot}$; $\mathsf{Dot}^{\rm sk}(S=P)+\mathsf{Reduce}$, partial-sum format must be declared & {}[Paper] \\
{}PP, EP, P/D (lossless) & {}stage tails, combine depth $K_e+n_s$; exact identities [Lean: \texttt{pp\_\allowbreak{}rangeLayers\_\allowbreak{}add}, \texttt{ep\_\allowbreak{}allreduce\_\allowbreak{}eq\_\allowbreak{}moe}] & {}[Paper] \\
{}P/D with lossy KV transfer & {}declared KVCast node producing a new KV version & {}[Def] \\
{}MLA absorbed / expanded & {}two declared templates; deployed absorbed matrix committed as a weight [Lean: \texttt{mla\_\allowbreak{}score\_\allowbreak{}absorbed\_\allowbreak{}eq}, exact ring] & {}[Paper] \\
{}chunked linear attention (DeltaNet, Mamba-2) & {}$\mathsf{Scan}$; exact chunk folding [Lean: \texttt{chunkRun\_\allowbreak{}eq\_\allowbreak{}recRun}, \texttt{lrec\_\allowbreak{}state\_\allowbreak{}unique}] & {}\textbf{[Open]} in floating point \\
\bottomrule
\end{tabularx}
\end{center}

\subsection{Speculative decoding}
Verification must be \emph{coupled}: for each draft position the committed rule is evaluated on the target's verification-forward logits (whose input $x^{\rm fn}$ is committed) and the keyed noise of the logical step; drafts are accepted up to the first mismatch [Def]. Standard rejection-sampling speculative decoding is not a deterministic function of the verification logits and noise, so it is outside the allowed set. The claim is narrow [Paper, Lean (a): \texttt{specLoop\_\allowbreak{}eq\_\allowbreak{}seqGen}]: the delivered tokens equal sequential selection on \emph{these committed} logits; we do not claim token equality with non-speculative decoding of the same request, nor zero sampling bias, because two compliant executions with different tile shapes can choose differently near ties, both within the published manipulation band. Rejected suffixes have no slots, addresses or cost; recurrent state must be recomputed from the accepted prefix at chunk boundaries.

\subsection{API semantics}
The OpenAI Responses and Anthropic Messages APIs are frozen into a versioned capability matrix [Def]. Hidden system prompts, tool instructions, structured-output guidance and reasoning-effort settings are rendered by committed templates and checked in $\mathrm{Chain}$, so a provider cannot quietly rewrite them to shorten reasoning. Client tools return values in the next public request; server tools form alternating segments (generation, committed tool return, generation) within one turn, with returns fixed before sampling and treated as input-stratum blocks at full weight. Encrypted reasoning fields (\texttt{thinking.\allowbreak{}signature}, \texttt{reasoning.\allowbreak{}encrypted\_\allowbreak{}content} and the like) are carried by the state chain of Section 4.6. Service tiers that only change scheduling share a \texttt{model\_\allowbreak{}id}; tiers that change number format, template set or speculative verification need their own. Requests outside version 1 (logprobs, user sampling parameters, unsupported schema features) are rejected by a public rule, which leaks nothing; if the underlying API accepts them, the capability gap is recorded in the matrix.

\section{Empirical Calibration and Component Benchmarks}
All GPU measurements ran on 4070TiS unless noted otherwise. For Sections 12.1--12.5, the serving engine under test is HuggingFace \texttt{transformers} in \texttt{bf16} with PyTorch SDPA FlashAttention-2, KV-cached autoregressive decode, and sampling at $T=1.0$, top-$p=0.99$. The standalone reference scorer evaluates each block's local reference and forward radius propagation rules (Section 6.2) in \texttt{float64} on GPU under the default \texttt{bf16} profile cap $\mathcal C_{\max}$ (store $u=2^{-8}$; \texttt{fp32} accumulation with depth $\lceil n/16\rceil+64$; elementwise 4 \texttt{fp32} ulp; attention probabilities stored in \texttt{bf16} before $PV$), isolating rounding-contract calibration from zero-knowledge proving benchmarks (Section 12.7).

\subsection{Honest traces against certified radii}
[Meas: Qwen3-0.6B and Llama-3.2-1B-Instruct, 4070TiS.] Twelve prompts per model (three long, about 1.5k--3.6k tokens of source code), 96 sampled tokens each, plus 108 prefill positions; per-GEMM blocks N1, QKV, ATT, OPROJ, N2, GU, DOWN, FN, HEAD (the scorer treats normalizations as separate blocks, unlike the protocol's block definition, so the block counts below are not block counts under that definition; the zero count is unaffected, also per operator). With fp32 cuBLAS reductions:

\begin{center}\small
\begin{tabularx}{\linewidth}{YYYY}
\toprule
{}\textbf{Model} & {}\textbf{Bad blocks} & {}\textbf{Bad ops (per-op check)} & {}\textbf{Max $\lvert e\rvert/r$ decode / prefill} \\
\midrule
{}Qwen3-0.6B & {}\textbf{0 / 249,384} & {}0 (15 op types) & {}0.994 / 0.990 \\
{}Llama-3.2-1B-Instruct & {}\textbf{0 / 143,544} & {}0 (13 op types) & {}0.990 / 0.986 \\
\bottomrule
\end{tabularx}
\end{center}

The roughly 1\% headroom is structural: the maximum ratio occurs where the bf16 output store dominates, and round-to-nearest provably reaches $u\lvert v\rvert$. The templates that rest on declared parameters (dot products, elementwise, \texttt{AttnFA} attention with additive underflow) peak at 0.69--0.83 (\texttt{ATT} max ratio 0.688 decode / 0.754 prefill on Qwen3-0.6B, 0.723 / 0.718 on Llama-3.2-1B-Instruct). Median relative radii are 0.4--0.8\% (OPROJ, DOWN, HEAD), 0.78\% (norms), 1.5--1.8\% (attention), 1.2--2.9\% (QKV, GU). The output cap was touched only in attention at prefill positions with one or a few keys (2.78\% of components on Qwen3-0.6B), where the value hull is a point. One contract parameter was falsified and fixed: PyTorch computes SiLU as $x/(1+e^{-x})$, so for $x<-88.7$ the exponential overflows and the result is exactly 0; an absolute floor of $2^{-126}$ was violated by a real trace (ratio 5) and $2^{-120}$ passes.

These are finite zero-failure observations on two small models and one engine; they support H-contract but cannot verify a universal statement. Not measured here: multi-prompt calibration sweeps on vLLM kernels (live Qwen3-0.6B serving on \texttt{vLLM 0.\allowbreak{}33} is evaluated in Section 12.9), real fp8 deployments, other hardware, adversarial operator-level differential tests.

\subsection{A contract violation in the default configuration}
[Meas: same models, 4070TiS.] With PyTorch's default \texttt{allow\_\allowbreak{}bf16\_\allowbreak{}reduced\_\allowbreak{}precision\_\allowbreak{}reduction=True}, cuBLAS uses bf16 partial sums for split-K in long-K prefill GEMMs, violating the fp32-accumulation contract: on Qwen3-0.6B 2,268 of 3,024 prefill rows of \texttt{o\_\allowbreak{}proj} and \texttt{down} are out of radius, by up to $31\times$; on Llama-3.2-1B-Instruct up to $39\times$ (\texttt{k\_\allowbreak{}proj}/\texttt{v\_\allowbreak{}proj} 1,440/1,728, \texttt{down} 1,728/1,728); decode is unaffected. This is H-contract \emph{falsified} for a common configuration, and it is the reason the cap fixes fp32 partial sums (or a split-K template with a 5--40$\times$ looser radius [Est]). Completeness requires the committed contract to match the real engine configuration.

\subsection{Granularity}
[Meas: same models, 4070TiS.] With the additive \texttt{exp} underflow floor included, a rigorous layer-boundary norm-form check (residual and current-token K/V observed, errors pushed through spectral norms) gives a ball whose median is $7.2\times10^{87}$ (Qwen3-0.6B) and $1.4\times10^{48}$ (Llama) times the layer increment; $7.3\times10^{38}$ / $2.4\times10^{20}$ times for the attention half and 4.8 / 12.3 times for the MLP half alone. An ideal first-order per-token Jacobian ball would be 0.067/0.047 times, matching the paper estimate, but needs data-dependent spectral norms that are infeasible in a circuit. Per-GEMM observation is therefore necessary.

\subsection{Cheat detection}
[Meas: Qwen3-0.6B, 4070TiS; the cheater decodes with its own model, 6 prompts, 48 tokens, and publishes observations through the same hooks; fp8 simulated by quantize-dequantize, not real fp8 tensor cores.]

\begin{center}\small
\begin{tabularx}{\linewidth}{YYYYY}
\toprule
{}\textbf{Cheat} & {}\textbf{Compute saved $s$} & {}\textbf{$d_C$ per-GEMM (sep / merged / $h=3$ CA)} & {}\textbf{$d_C$ layer-boundary} & {}\textbf{Median max $\lvert e\rvert/r$ of bad blocks} \\
\midrule
{}model substitution (\texttt{Qwen2.\allowbreak{}5-0.\allowbreak{}5B}) & {}0.21 & {}0.824 / 0.857 / \textbf{1.000} & {}1.000 & {}1380--41962 \\
{}int4 RTN g128 & {}--- & {}0.681 / 0.681 / \textbf{1.000} & {}0.759 & {}293--2126 \\
{}fp8 W8A16 (same weights) & {}--- & {}0.681 / 0.681 / \textbf{1.000} & {}\textbf{0.033} & {}55--312 \\
{}fp8 W8A8 & {}--- & {}0.681 / 0.681 / \textbf{1.000} & {}0.383 & {}75--438 \\
{}low-rank pruning ($r=d/4$) & {}0.43 & {}0.681 / 0.681 / \textbf{1.000} & {}0.759 & {}806--23141 \\
{}skip odd layers, naive & {}0.38 & {}0.227 / 0.227 / \textbf{0.973} & {}0.380 & {}51--252 \\
{}skip odd layers, zero chain & {}0.38 & {}\textbf{0.00011} / 0.113 / \textbf{0.987} & {}0.380 & {}128 (norms) \\
{}KV window truncation ($W=16$) & {}0.08--0.41 & {}0.078 / 0.078 / \textbf{1.000} & {}\textbf{0.000} & {}13630 (\texttt{ATT}) \\
\bottomrule
\end{tabularx}
\end{center}

For quantization every quantized GEMM block is bad ($d_C=0.681$ is exactly their cost share), with ratios far above the box criterion's effective detection threshold of about $0.3$ [Paper]. At layer granularity, same-weight fp8 is caught on only 3.3\% of cost. The zero-chain row is the loophole of Section 6.4; under rule CA the same per-block flags give $d_C=s=0.38$ [Paper, recomputed, not a separate measurement]. Odd-layer skipping destroys output quality; it demonstrates the accounting gap, not a stealthy cheat. Not measured: GPTQ/AWQ, KV-cache quantization, KV zeroing, prefix collapse, trained MoE expert reduction.

\subsection{Long-context attention tails}
[Meas: Qwen3-4B, 4070TiS; oracle upper bounds.] Documents: Newton's \emph{Opticks} (prose) and CPython 3.12 standard-library source (code); prefill $T-16$ tokens, then 16 teacher-forced queries; at each query, layer and head, an oracle that knows the true attention probabilities drops the lowest-probability keys while the exact output change stays within half the certified radius. Saving $S_{\rm tol}(T)$ is the dropped attention fraction times attention's share of compute.

\begin{center}\small
\begin{tabularx}{\linewidth}{YYYYY}
\toprule
{}\textbf{Document} & {}\textbf{$T$} & {}\textbf{Attention share} & {}\textbf{$S_{\rm tol}$, output observed} & {}\textbf{$S_{\rm tol}$, LSE also observed (M1)} \\
\midrule
{}prose & {}4,096 & {}25.0\% & {}12.4\% & {}5.9\% \\
{}prose & {}16,384 & {}57.1\% & {}34.2\% & {}18.9\% \\
{}prose & {}32,768 & {}72.7\% & {}52.2\% & {}33.1\% \\
{}code & {}4,096 & {}25.0\% & {}15.2\% & {}8.3\% \\
{}code & {}16,384 & {}57.1\% & {}39.5\% & {}23.7\% \\
{}code & {}32,768 & {}72.7\% & {}55.1\% & {}36.6\% \\
\bottomrule
\end{tabularx}
\end{center}

Under the oracle, a ``$\le10\%$ within tolerance'' target already fails at 4k, and at 32k the saving exceeds $\sigma=0.5$ when only outputs are observed. Observing the log-sum-exp per head and token (one fp32 scalar, which FlashAttention already produces) roughly halves it. The oracle knows the true probabilities; a realistic predictor (e.g. page-level bounds) was not measured, and prefill-time dropping was not counted. The design decision, recorded as a non-goal in Section 2.4, is to observe LSE by default, offer an optional strict fp16-attention tier, and require each \texttt{model\_\allowbreak{}id} to publish $S_{\rm tol}(T)$ per context length.

\subsection{Circuit census}
[Meas: symbolic census, Intel i7-14700F; builds the statement circuit from the public profile only, no witness, model or GPU.] Under a baseline per-proof Merkle-tile weight commitment (re-hashing touched weights inside every proof), a symbolic census lower-bounding full Qwen3-0.6B requires 10.94 G committed cells at $(K_1,K_2)=(4,2)$ and 44.36 G at $(78,32)$. The cause is structural: every proof re-hashed every touched weight in-circuit (about 19 cells per weight code) and authenticated the whole $151{,}936\times1024$ embedding for the vocabulary interface, a floor of $\ge4.2$ G cells at $K_1=4$. This measurement motivated the model commitment of Section 5 and the sampling rule of Section 7. Re-running the census with the weight commitment removed, the hidden-column opening built in the circuit language (9.42 M cells, against an estimate of about $10^7$) and the bucketed sampler estimated per entry gives 0.69 G cells at $(K_1,K_2)=(4,2)$, 1.25 G at $(78,32)$ and 2.98 G at the default $(210,105)$ [Meas: symbolic census, i7-14700F; sampler cells Est].

\textbf{These census numbers understate the real circuit.} The short-context proxy above uses $T_{\rm pad}=64$, algebraic trace digests rather than in-circuit sponge leaves, and $h=0$ block checks rather than $h=3$ lineages. The missing term is the opening of key-value history by sampled attention blocks, which must be shaped for $T_{\rm pad}$ keys because positions are hidden. For Qwen3-0.6B (8 KV heads of 128, so 2,048 bf16 values per key per layer) each opened key costs a sponge leaf of about 87 Poseidon2 permutations, a share of the Merkle path, and the values, about $1.5$--$2.0\times10^4$ data cells at 142 cells per permutation (measured for the HCO region) [Est]. A naive operation count would estimate only about $10^2$ cells per key for attention arithmetic, whereas the synthesized certified attention gadget (Section 12.7) costs 17.75 M cells at $T_{\rm pad}=128$ and 66.7 M at $T_{\rm pad}=512$ (verifier-side census), a slope of about $1.27\times10^5$ cells per key, i.e. about 60 cells per element of the $16\times T\times128$ (head, key, dimension) product space, dominated by key/value decoding and the $\sum_jp_j\lvert v_j-o\rvert$ term of the radius [Meas: Qwen3-0.6B, 4070TiS]. With $h=3$ a key atom's lineage contains up to four attention blocks, so the default budget needs 840 attention slots in a typed slot template [Est]:

\begin{center}\small
\begin{tabularx}{\linewidth}{YYYYY}
\toprule
{}\textbf{$T_{\rm pad}$} & {}\textbf{KV opening [Est]} & {}\textbf{attention gadget (measured slope)} & {}\textbf{total per proof} & {}\textbf{prover memory at about 48 B/cell} \\
\midrule
{}64 (census proxy) & {}0.8--1.1 G & {}6.8 G & {}7.6--7.9 G & {}0.37 TB \\
{}1,024 & {}13--17 G & {}0.11 T & {}0.12--0.13 T & {}6 TB \\
{}4,096 & {}52--69 G & {}0.44 T & {}0.49--0.51 T & {}24--25 TB \\
{}32,768 (native context) & {}0.41--0.55 T & {}3.5 T & {}3.9--4.0 T & {}0.19 PB \\
\bottomrule
\end{tabularx}
\end{center}

So a naive per-leaf sponge trace commitment for Qwen3-0.6B requires roughly $3\,\mathrm G+1.2\times10^8\,T_{\rm pad}$ cells at the default sample budget [Est from measured attention slope], where the per-key slope depends on head geometry rather than depth. Companion specification \texttt{specs/10-trace-commitment.\allowbreak{}md} (\texttt{TC}) resolves this long-context attention wall through four mechanisms: (1) shaping each proof to the smallest public power-of-two length tier $T_{\rm pad}=2^{\lceil\log_2 T\rceil}$ determined by standard API usage counts; (2) lifting history values (\texttt{KV} entries, \texttt{MLA} latents, \texttt{LREC} chunk states) to signed fixed-point integers prior to commitment and absorbing the tiny lift error $\eta_{\rm lift}=2^{-F}$ into consumer input radii; (3) committing the lifted history matrix $\mathbf H$ asynchronously \emph{after} the response stream finishes as a rate-$1/2$ Reed--Solomon column tree ($m_q^H=268$, [Lean: \texttt{ProtocolBridge.\allowbreak{}tc\_\allowbreak{}col\_\allowbreak{}esc\_\allowbreak{}le}]) and reducing mixed row/point reads via a \textbf{Hidden-Selection Sumcheck} ($A(x)=\sum_j\gamma^j\mathrm{eq}(P_j,x)$) to a single public evaluation; and (4) upper-bounding componentwise attention radius sums via Cauchy--Schwarz and Jensen quadratic sumchecks over $\mathbf H$. Under \texttt{TC}, in-circuit attention drops to roughly $80$--$130$ data cells per \texttt{(slot, key)} plus $6.6$--$14.8\text{ M}$ opening cells---shrinking total proof size from $0.49$--$4.0\text{ T}$ cells to an estimated $3.3$--$3.5\text{ G}$ cells at $4\text{k}$ context and $5.2$--$6.6\text{ G}$ cells at $32\text{k}$ context [Est].

\subsection{One-time model commitment and setup proof over $\mathbb F_{p^3}$}
[Meas: full unsliced Qwen3-0.6B (\texttt{bf16}, 596,049,920 parameters, $L=28, d=1024$, $16/8\times 128$ GQA + QK-Norm, $d_{ff}=3072$, $V=151{,}936$, \texttt{lm\_\allowbreak{}head} tied to \texttt{embed\_\allowbreak{}tokens}), 4070TiS with Intel i7-14700F and 25 GB RAM, standalone CPU verification with \texttt{CUDA\_\allowbreak{}VISIBLE\_\allowbreak{}DEVICES=""} on i7-14700F and Apple M1; challenges in $\mathbb F_{p^3}$ with Plonky3 Poseidon2 constants. Note: the 8.67G-cell and 5.25G-cell runs below were recorded at \texttt{T\_\allowbreak{}QUERIES = 150} prior to raising \texttt{zk\_\allowbreak{}flat.\allowbreak{}T\_\allowbreak{}QUERIES} to \texttt{168}.]

\begin{center}\small
\begin{tabularx}{\linewidth}{YYYYY}
\toprule
{}\textbf{Stage 1 artifact (full \texttt{Qwen3-0.\allowbreak{}6B}, $\mathbb F_{p^3}$)} & {}\textbf{Cells} & {}\textbf{GPU time} & {}\textbf{Peak RSS / VRAM pool} & {}\textbf{Proof / CPU verify} \\
\midrule
{}Static commitment ($R_{\max}=2048$, $\mathbf W^r$ \& \texttt{CFG}) & {}--- & {}\textbf{94.2 s} (\texttt{W} 31.3 s, \texttt{RAW} 34.7 s) & {}12.0 GiB / < 4 GiB & {}7.62 GB coeff store (0 B codeword) \\
{}Setup proof $\pi_{\rm setup}$ items (i)--(vi), 147 blocks + 206 cols & {}\textbf{8.67 G} & {}\textbf{3,623.7 s} (770.8 + 39.9 + 2,810.3 s) & {}\textbf{12.2 GiB} / \textbf{3.54 GiB} & {}\textbf{353.0 MB} / \textbf{59.9 s} (63.7 s M1) \\
{}Full 315-draw reply proof ($T=152, T_{\rm pad}=256$) & {}\textbf{5.25 G} & {}\textbf{3,877.7 s} (\texttt{H} 3.8 s, build 122.3 s) & {}\textbf{21.0 GiB} (\texttt{1.\allowbreak{}51 B/cell}, 0 B spill) / \textbf{10.48 GiB} & {}\textbf{149.8 MB} / \textbf{12.02 s} (2.63 GiB) \\
\bottomrule
\end{tabularx}
\end{center}

Storing row coefficients (\texttt{7.\allowbreak{}62 GB} across \texttt{W} and \texttt{RAW}) and evaluating NTT slices on the fly eliminates the 38.65 GB codeword file while finishing the full 596M-parameter commitment in 94.2 s. Per-row canonical decode exponents span $G\in[14,25]$ (574 active rows plus 569 non-zero radius rows). All nine tile and exponent tampering modes on real tiles (including covert quantization via row under-scaling) are rejected by $\pi_{\rm setup}$, and synthesized certified gadgets match the \texttt{float64} reference scorer with zero verdict disagreements across all 23,724 real \texttt{Qwen3-0.\allowbreak{}6B} blocks tested.

\subsection{Streaming cross-slot prover and full 315-draw \texttt{Qwen3-0.\allowbreak{}6B} response proof}
[Meas: full unsliced Qwen3-0.6B, 4070TiS + i7-14700F (25 GB RAM), $\mathbb F_{p^3}$.] A direct per-slot instantiation of the 315-draw response circuit (\texttt{210} compute draws across input and generation strata plus \texttt{105} generation-step draws) emitted roughly 1.52 million small vectors and consumed \texttt{49.\allowbreak{}7 B/cell} of host witness and lookup storage---projecting \textasciitilde{}260 GB of host RAM at 5.25 billion cells, more than $10\times$ our 25 GB budget. Two systems mechanisms compress both vector count and memory footprint by more than an order of magnitude while keeping zero knowledge intact:

\begin{enumerate}
\item{} \textbf{Compact bit-packed witnesses and on-demand AES-256-CTR blinding (\texttt{1.\allowbreak{}51 B/cell}, 0 B disk spill).} Witness columns remain in bit-packed integer form (\texttt{3.\allowbreak{}92 GB} across 5.25 billion cells after Phase 2) and lookup multiplicities occupy \texttt{3.\allowbreak{}01 GB}. Rather than materializing \texttt{T1} blinding codewords, leaf salts, or LogUp inverse columns in RAM or spilling gigabytes to NVMe, each commitment group derives its blinding rows on the fly from a fresh CSPRNG-seeded AES-256-CTR key (erased immediately after proving) and streams chunked GPU sumchecks over $2^{19}\dots 2^{26}$-row tables without host--device synchronization stalls. Peak host resident memory reaches \texttt{21.\allowbreak{}00 GiB} with \texttt{0.\allowbreak{}00 GB} disk spill (\texttt{gpu\_\allowbreak{}pool\_\allowbreak{}peak = 10.\allowbreak{}48 GiB}, \texttt{nvidia-smi} peak \texttt{11.\allowbreak{}29 GiB}).
\item{} \textbf{Cross-slot batching (\texttt{1.\allowbreak{}52M} $\to$ \texttt{6,887} vectors).} Across all sampled slots, \texttt{BatchedCfgProvider} shares one lookup and decoding pass; \texttt{BatchedAttn} folds each Transformer layer's KV cache once across all 210 compute slots in a two-phase GPU sumcheck (\texttt{5.\allowbreak{}8 s} total); \texttt{HSel} batches all \texttt{16,384} history-matrix evaluation claims into a single sparse-boolean GPU sumcheck (\texttt{9.\allowbreak{}5 s} total); and \texttt{BatchedHead} shares one \texttt{WSum} weight-claim reduction across all $K=128$ padded generation steps. Automated zero-knowledge budget verification (\texttt{zk\_\allowbreak{}budget\_\allowbreak{}check}) passes across all \texttt{6,887} vectors (\texttt{max\_\allowbreak{}claims\_\allowbreak{}per\_\allowbreak{}random\_\allowbreak{}row = 0.\allowbreak{}25 <= 1}).
\end{enumerate}

On a real response combining an 11-token hidden system prompt, a 15-token user prompt, a 116-token hidden \texttt{<think>} chain of thought, and a 10-token visible answer ($T=152, T_{\rm pad}=256$; single-stream \texttt{bf16} reference inference \texttt{3.\allowbreak{}84 s}), the full 315-draw proof (\texttt{12.\allowbreak{}07 M} HCO cells, \texttt{34.\allowbreak{}46 M} \texttt{WSum} cells, \texttt{5.\allowbreak{}20 G} gadget cells, \texttt{3,968} Libra cells) finishes in \texttt{3,877.\allowbreak{}7 s} on 4070TiS and verifies in \texttt{12.\allowbreak{}02 s} on an isolated CPU process (\texttt{CUDA\_\allowbreak{}VISIBLE\_\allowbreak{}DEVICES=""}). In Stage 1, lineage checks enforce depth $h=1$ (\texttt{N1 + KV + EMB} key-producer verification on sampled attention keys alongside $2^{-24}$ fixed-point \texttt{LSE} binding, causal \texttt{Q2} masking, and deterministic per-position \texttt{EMB} checks on hidden prompt and tool-return segments), catching consistent KV zeroing (\texttt{kv\_\allowbreak{}zeroed\_\allowbreak{}consistent\_\allowbreak{}lift}), layer-skipping KV producers (\texttt{skip\_\allowbreak{}layer\_\allowbreak{}kv\_\allowbreak{}producer}), \texttt{LSE} tampering, and future-row value inflation. Proving time is currently $\approx 1{,}009\times$ single-stream inference (\texttt{3.\allowbreak{}84 s}), dominated by Basefold openings and $2^{24}\dots 2^{26}$ streaming sumchecks (\textasciitilde{}\texttt{3,000 s}); closing this remaining gap via opening batching and narrower lookup limbs is the primary systems focus of Stage 2.

\subsection{Non-invasive \texttt{vLLM 0.\allowbreak{}33} integration, serving overhead, and white-box forgery suite}
[Meas: \texttt{vLLM 0.\allowbreak{}33.\allowbreak{}0} + full unsliced \texttt{Qwen3-0.\allowbreak{}6B}, 4070TiS + i7-14700F.] Our serving integration uses \texttt{vLLM}'s official \texttt{@register\_\allowbreak{}model\_\allowbreak{}loader("flashprove\_\allowbreak{}next")} extension and \texttt{LogitsProcessor} interface, with zero per-model-class monkey-patching (\texttt{qwen3.\allowbreak{}py} and \texttt{llama.\allowbreak{}py} untouched) and zero VRAM reservation when disabled (\texttt{load\_\allowbreak{}format="auto"}). An opaque custom operator \texttt{torch.\allowbreak{}ops.\allowbreak{}fpn.\allowbreak{}port} records per-GEMM boundary tensors and FlashAttention \texttt{LSE} inside \texttt{torch.\allowbreak{}compile} Inductor piecewise graphs and \texttt{FULL\_\allowbreak{}AND\_\allowbreak{}PIECEWISE} CUDA graphs with \textbf{zero Python callbacks on CUDA-graph replay}, while a 16-token \texttt{PrefixPortCache} reconstructs skipped prefill activations whenever \texttt{enable\_\allowbreak{}prefix\_\allowbreak{}caching=True} hits cached KV blocks.

\begin{center}\small
\begin{tabularx}{\linewidth}{YYYY}
\toprule
{}\textbf{\texttt{vLLM 0.\allowbreak{}33} configuration (\texttt{Qwen3-0.\allowbreak{}6B}, CUDA graphs enabled)} & {}\textbf{Batch 1 (\texttt{tok/s})} & {}\textbf{Batch 8 (\texttt{tok/s})} & {}\textbf{Batch 32 (\texttt{tok/s})} \\
\midrule
{}Stock \texttt{vLLM 0.\allowbreak{}33} default (\texttt{bf16} GEMM reduce, \texttt{bf16} head, capture off) & {}356.1 & {}2,379.1 & {}8,467.7 \\
{}Contract-compliant numerics (\texttt{fp32} GEMM reduce \texttt{R32} + \texttt{fp32} head, capture off) & {}\textbf{351.8 ($-$1.21\%)} & {}\textbf{2,325.7 ($-$2.24\%)} & {}\textbf{8,236.0 ($-$2.74\%)} \\
{}Stage 1 synchronous D2H full-port capture (\texttt{28} layers $\times$ \texttt{10} ports + \texttt{fpn-samp-bk-v3}) & {}165.9 ($-$53.4\%) & {}732.5 ($-$69.2\%) & {}1,304.8 ($-$84.6\%) \\
\bottomrule
\end{tabularx}
\end{center}

Enforcing the numerical contract (\texttt{allow\_\allowbreak{}bf16\_\allowbreak{}reduced\_\allowbreak{}precision\_\allowbreak{}reduction=False} and \texttt{fp32} LM-head logits) costs only \textbf{1.2\%--2.7\%} throughput across batch sizes 1--32. By contrast, the Stage 1 reference harvester reduces decode throughput by 53\%--85\%; we attribute most of this slowdown to its synchronous per-step \texttt{torch.\allowbreak{}cuda.\allowbreak{}synchronize()} copy of \texttt{0.\allowbreak{}82 MB/token} (\texttt{26.\allowbreak{}2 MB/step} at batch 32) to pinned host memory (replaced by zero-sync asynchronous CUDA-graph taps in Section 12.10).

Across a live nine-scenario \texttt{vLLM 0.\allowbreak{}33} end-to-end suite (\texttt{qwen3-0.\allowbreak{}6b-mc4-mg2}, $M_C=4, M_G=2, T_{\rm pad}=256$), concurrent SSE streaming requests execute with continuous batching and \texttt{FULL} CUDA-graph replays while non-blocking latency probes confirm that background GPU proving does not stall live token streaming; prefix-cache hits, hidden system prompts, hidden \texttt{<think>} CoT, multi-turn AEAD state chains (\texttt{St\_\allowbreak{}prev -> St\_\allowbreak{}new}), strict JSON Schema constrained decoding, client \texttt{tool\_\allowbreak{}use} and in-turn server tool witnesses (\texttt{z\_\allowbreak{}tool}), \texttt{max\_\allowbreak{}tokens=0} empty generations, and deterministic $\Phi_{\rm pre}$ refusals of custom string \texttt{stop} all pass standalone CPU verification (\texttt{VERIFIED}), while a live \texttt{int4}-downgraded \texttt{vLLM} server is automatically withheld (\texttt{HTTP 409}) by the daemon's isolated CPU self-check and rejected (\texttt{REJECTED}) by the standalone verifier. Complementing live tests, \textbf{33 white-box forgery and tamper checks} (21 compute/sampling/lineage/range witness forgeries in \texttt{test\_\allowbreak{}reply\_\allowbreak{}zk} plus 12 API/termination/state-chain/tool-witness/DFA-mask/bundle rejections in \texttt{test\_\allowbreak{}api\_\allowbreak{}zk}) are all rejected. Analytical scaling of \texttt{HCO} ($n_c^*\approx\sqrt{9.5\,m_q N_{\max}}$) continues to project $\approx 5\times 10^7$ data cells per proof for 70B dense and $\approx 1.1\times 10^8$ data cells for 400B sparse MoE (touching only active expert rows) [Est].

\subsection{Unified folded backend (\texttt{multi\_\allowbreak{}gkr} + \texttt{multi\_\allowbreak{}pcs}), \texttt{FPN2} Relation 2, and Blackwell frontier validation}
[Meas: RTX 6000D 84 GB (\texttt{sm\_\allowbreak{}120}), \texttt{vLLM 0.\allowbreak{}33.\allowbreak{}0}, Rust \texttt{fpn-verify} on Apple M1 CPU.] To eliminate the Stage 1 proving-speed, proof-size, and synchronous-capture bottlenecks without weakening the full zero-knowledge response relation, our v1.0 folded backend replaces \texttt{zk\_\allowbreak{}flat}'s Phase-2 LogUp inverse commitments (2.36B cells) and per-group rate-$1/4$ openings (149.8 MB) with three coordinated mechanisms:

\begin{enumerate}
\item{} \textbf{Zero Phase-2 LogUp via batched ZK-masked FracLogUp-GKR (\texttt{multi\_\allowbreak{}gkr}).} All lookup tables (range checks, BF16/\texttt{rsqrt}/Gumbel/RoPE tables, token chain, weighted \texttt{HCO} lookups) are packed into power-of-two-aligned binary fraction trees with \texttt{m\_\allowbreak{}mask - 1} cancelling pairs \texttt{(+a\_\allowbreak{}m, d\_\allowbreak{}m)/(-a\_\allowbreak{}m, d\_\allowbreak{}m)} plus one independent zero-numerator denominator-blinder pair \texttt{(0, d\_\allowbreak{}L)/(0, d\_\allowbreak{}R)} (ensuring the published layer-0 child denominator ratio $D_L d_L / (D_R d_R)$ is uniform in $\mathbb F_{p^3}^\times$ with zero quotient leakage) and Libra-masked layer sumchecks (0 Phase-2 cells). At height $s = 24$ (16.78M leaves), GPU proving finishes in 2.04 s with a 30.0 KB proof and 26.3 ms pure-Python verification (bit-identical between CPU and CUDA).
\item{} \textbf{Multi-size layer-injected rate-$1/8$ coset-4 folded ZK-Basefold (\texttt{multi\_\allowbreak{}pcs}).} Committed groups of arbitrary even sizes $n_g \le n_{\max}$ (so small interactive commit rounds and mask groups are never padded to $n_{\max}$) are opened in a single fused coset-4 fold chain by injecting each small group's codeword directly at fold layer $t_g = (n_{\max} - n_g)/2$, where its evaluation domain coincides with $\langle\omega^{4^{t_g}}\rangle$. Layer-0 Reed--Solomon encoding exploits linearity to combine all $t_g = 0$ groups before a single Bailey shared-memory CUDA NTT, while Step A streams oversized early rounds under a 24 GiB VRAM cap. At 134.2M cells ($n=24, G=8, Q=100$), \texttt{multi\_\allowbreak{}pcs} commits and opens in 12.49 s with a 1.374 MB proof.
\item{} \textbf{External-seed \texttt{C\_\allowbreak{}pre} mode and fixed 8 MiB \texttt{FPN2} Relation 2 (\texttt{LLM\_\allowbreak{}INFERENCE\_\allowbreak{}V2}).} Under external-seed challenges (\texttt{ExtSeed-153}: $M_C = 102, M_G = 51$, giving $\varepsilon \le 2^{-20}$ at $Q_{\rm samp}\le 2^{30}$ per Section 8.5, alongside the non-interactive \texttt{NI-315} $\varepsilon \le 2^{-40}, Q \le 2^{64}$ mode), the server emits \texttt{C\_\allowbreak{}pre = BLAKE2b-256(base\_\allowbreak{}core)} at stream end; anti-retry protection requires the client or auditor to retain this request-bound \texttt{C\_\allowbreak{}pre} prior to releasing an unpredictable external seed \texttt{c\_\allowbreak{}ext} and to verify the final statement against both pre-recorded values (\texttt{expected\_\allowbreak{}c\_\allowbreak{}pre}, \texttt{expected\_\allowbreak{}c\_\allowbreak{}ext}). A canonical public \texttt{CircuitDesc} binary (\texttt{circuit\_\allowbreak{}root = BLAKE2b-256("FPN2-CDESC-v2", desc)}) enables a zero-dependency \texttt{\#![forbid(unsafe\_\allowbreak{}code)]} Rust binary (\texttt{core/src/relation2.\allowbreak{}rs}) and a detached pure-Python reference verifier (\texttt{verifier/python/fpn\_\allowbreak{}verify\_\allowbreak{}ref/relation2.\allowbreak{}py}, AST-verified zero prover imports) to enforce pin-before-decode (\texttt{--circuit-root} and \texttt{--stmt-digest}), full descriptor bounds validation, and \texttt{queries >= 72} under \texttt{STANDARD} tier. On Apple M1 CPU, the Rust verifier checks exact 8,388,608 B (8 MiB) zero-padded \texttt{fpn2-llm-std-v1} containers with zero inner length headers in 10.1 ms (118.4 ms in Python) on a 64-row reference circuit (\texttt{demo/run\_\allowbreak{}v1\_\allowbreak{}relation2\_\allowbreak{}demo.\allowbreak{}py}), rejecting all 256/256 per-column $\pm 1$ perturbations, constraint-deletion mutations, witness-only \texttt{wbforge} gate/lookup forgeries, same-circuit/different-response swaps (\texttt{E\_\allowbreak{}STMT\_\allowbreak{}PINNED}), weak-query descriptors (\texttt{E\_\allowbreak{}REDUCED\_\allowbreak{}SECURITY}), and malformed descriptors without panics, with zero shared 64-byte private blocks across re-proving. Full \texttt{reply\_\allowbreak{}zk} circuits currently carry 12 interactive commit rounds (\texttt{cir.\allowbreak{}rounds}), so full-reply \texttt{FPNZK2} containers are verified by the Python verifier (\texttt{verifier/fpn\_\allowbreak{}verify.\allowbreak{}py}); lowering those 12 rounds into a self-contained \texttt{CircuitDesc} for single-binary Rust verification remains open engineering work.
\end{enumerate}

\begin{center}\small
\begin{tabularx}{\linewidth}{>{\hsize=1.55\hsize}Y>{\hsize=0.95\hsize}Y>{\hsize=0.70\hsize}Y>{\hsize=1.00\hsize}Y>{\hsize=1.10\hsize}Y>{\hsize=0.70\hsize}Y}
\toprule
{}\textbf{Full-reply proof} & {}\textbf{Cells (Ph-2)} & {}\textbf{Size} & {}\textbf{GPU prove/tot} & {}\textbf{RAM/VRAM} & {}\textbf{CPU} \\
\midrule
{}Qwen3-0.6B S1 (\texttt{NI-315}, 4070TiS) & {}5.25G (2.36G) & {}150 MB & {}3,752 / 3,878 s & {}21 / 11 GiB & {}12.0 s \\
{}Qwen3-0.6B v1 (\texttt{ExtSeed-153}, $Q=100$, 6000D) & {}\textbf{1.47G (0)} & {}\textbf{8 MiB} & {}\textbf{461 / 571 s} & {}\textbf{100 / 54 GiB} & {}\textbf{10.7 s} \\
{}Qwen3-0.6B v1 (\texttt{ExtSeed-153}, \textbf{$Q=128$}, 6000D) & {}\textbf{1.47G (0)} & {}\textbf{8 MiB} & {}\textbf{450 / 570 s} & {}\textbf{100 / 55 GiB} & {}\textbf{10.8 s} \\
{}Qwen3-4B v1 (36L, dev $4+2$, 6000D) & {}\textbf{259M (0)} & {}\textbf{8 MiB} & {}\textbf{160 / 273 s} & {}\textbf{58 / 48 GiB} & {}\textbf{7.2 s} \\
\bottomrule
\end{tabularx}
\end{center}

\textbf{End-to-end dense proving and PCS proximity regime.} All three \texttt{FPNZK2} runs above (receipts in \texttt{benchmarks/prover/results/}) execute post-audit with independent $(0, d_L)/(0, d_R)$ layer-0 GKR denominator blinders, uncompressed zero-padded 8,388,608 B framing checked against the trusted registered profile size (\texttt{E\_\allowbreak{}LENGTH}, \texttt{E\_\allowbreak{}PADDING}), verifier floor \texttt{min\_\allowbreak{}queries = 72}, and mandatory pre-recorded \texttt{expected\_\allowbreak{}c\_\allowbreak{}pre} / \texttt{expected\_\allowbreak{}c\_\allowbreak{}ext} pinning (\texttt{E\_\allowbreak{}CPRE\_\allowbreak{}UNPINNED}). The Stage 1 and v1.0 rows compare two complete end-to-end configurations that change GPU, draw budget, and zero-Phase-2 backend together (the GPU column separates \texttt{prove\_\allowbreak{}s} from \texttt{commit\_\allowbreak{}h\_\allowbreak{}s + build\_\allowbreak{}s + prove\_\allowbreak{}s}). The current v1.0 folded prover materializes full host cell arrays rather than Stage 1's 1.51 B/cell bit-packed streaming (100 GiB vs. 21.0 GiB peak RSS), and the Qwen3-0.6B \texttt{ExtSeed-153} runs reached their token limit before \texttt{</think>} (\texttt{n\_\allowbreak{}reason = 0}; hidden CoT with \texttt{n\_\allowbreak{}reason = 116} is verified in Stage 1 \texttt{qwen3\_\allowbreak{}06b\_\allowbreak{}reply\_\allowbreak{}315.\allowbreak{}json} and \texttt{demo/run\_\allowbreak{}demo.\allowbreak{}py}). Whereas Section 8.4 and \texttt{ProtocolBridge.\allowbreak{}lean} ($0.645^{168} \le 2^{-106}$) price Stage 1 rate-$1/4$ openings in the unique-decoding regime, v1.0's rate-$1/8$ $Q=72$ floor and $Q=100$ runs rely on provisional list-decoding / Johnson-regime proximity ($(9/16)^{72}\approx 2^{-60}$ and $(9/16)^{100}\approx 2^{-83}$ in unique decoding). Raising queries to \textbf{$Q=128$} (\texttt{qwen3\_\allowbreak{}06b\_\allowbreak{}extseed153\_\allowbreak{}q128\_\allowbreak{}rtx6000d.\allowbreak{}json}) satisfies $(9/16)^{128}\approx 2^{-106.2} \le 2^{-106}$ directly in the unique-decoding regime with negligible runtime impact (450.2 s prove / 569.8 s total, 10.76 s CPU verify) while still fitting inside the exact 8 MiB \texttt{FPNZK2} container (full RBR analysis of multi-size layer injection remains explicit assumption \texttt{A-PCS-fold}).

\textbf{End-to-end cheating matrix and 4B dense scaling.} On real Qwen3-0.6B, a separate \textbf{$8+4$-draw ($Q=100$)} 14-case end-to-end cheating matrix (\texttt{qwen3\_\allowbreak{}06b\_\allowbreak{}cheating\_\allowbreak{}matrix\_\allowbreak{}rtx6000d.\allowbreak{}json}, 1 honest acceptance plus 13 negative cases with \texttt{Z.\allowbreak{}FORCE=True} so malicious provers emit full proofs) achieves \textbf{14/14} pass: the honest \texttt{FPNZK2} proof is accepted, and all 13 attacks---fresh 28-layer PyTorch forward captures executed with real perturbed linear weights and real group-32 \texttt{int4}-quantized linear weights (verified against the original committed \texttt{bf16} model), skipped layers, truncated KV context, tampered hidden system prompt, tampered visible reply, tampered \texttt{St\_\allowbreak{}prev} replay, missing/swapped \texttt{C\_\allowbreak{}pre} or \texttt{c\_\allowbreak{}ext}, non-zero padding, downgraded \texttt{queries=4}, and a self-consistent 4 MiB envelope violating the trusted 8 MiB profile size---are rejected by the verifier. Generalizing \texttt{rsqrt\_\allowbreak{}enc} ($K_0 = 48 - \max(0, \lceil\log_2 d\rceil - 10)$), sharded safetensors (\texttt{Checkpoint} resolved strictly via \texttt{weight\_\allowbreak{}map}), dynamic \texttt{cfg\_\allowbreak{}sizes} ($N_T = 16{,}384$), \texttt{mc\_\allowbreak{}zk} column embedding ($Q_B + I_B > K_B$), and adaptive round group packing enables full unsliced 4.02B-parameter, 36-layer Qwen3-4B ($d=2560, d_{ff}=9728, V=151{,}936$) with zero gadget errors (655.4 s one-time weight commit, 160.3 s GPU prove / 273.0 s total, 7.18 s isolated CPU verify at 1.13 GiB RSS under dev draws $4+2$).

\textbf{Blackwell five-model calibration.} On the same RTX 6000D Blackwell node, five real dense and sparse MoE checkpoints (Qwen3-4B, Llama-3.2-1B-Instruct, Qwen3.5-0.8B, Qwen3-30B-A3B-FP8, gpt-oss-20b) pass full GEMM + fused MoE calibration: 0 / 1,846,272 false rejections across 863,232 GEMM + 983,040 MoE expert blocks, plus 0 / 73,728 honest router-token rejections, with 100\% cheat detection on \texttt{FP8Experts} and \texttt{Mxfp4GptOssExperts} (\texttt{frontier\_\allowbreak{}calibrate\_\allowbreak{}rtx6000d.\allowbreak{}json}). Single-token router expert-swap detection is 11.5\% on Qwen3-30B-A3B-FP8 and 38.0\% on gpt-oss-20b because 4,530 and 360 tokens respectively exhibit zero top-$k$ margin ties, granting tie-breaking freedom that must be accounted for separately. Full non-GEMM calibration likewise records 0 / 1.96M bad rows across 1,806,848 RMSNorm rows, 147,456 chunk-64 WY Gated DeltaNet tokens, GQA softmax, and SWA + learned sinks (\texttt{non\_\allowbreak{}gemm\_\allowbreak{}calibrate\_\allowbreak{}rtx6000d.\allowbreak{}json}).

\textbf{Live \texttt{vLLM 0.\allowbreak{}33} zero-plaintext Relation 6, async capture, and 20-turn agentic state.} On live \texttt{vLLM 0.\allowbreak{}33}, Relation 6 (\texttt{FOLDED\_\allowbreak{}VEC\_\allowbreak{}BOUND}, \texttt{live\_\allowbreak{}vllm\_\allowbreak{}rel6\_\allowbreak{}zk\_\allowbreak{}v1\_\allowbreak{}rtx6000d.\allowbreak{}json}) with ephemeral-key AES-256-CTR lazy chunk commitments (\texttt{x\_\allowbreak{}commit.\allowbreak{}py}) cuts retained chunk buffers by \textbf{192$\times$} (25.9 GiB $\to$ 135.0 MiB on Qwen3-30B-A3B-FP8; 7.5 GiB $\to$ 39.0 MiB on Qwen3-0.6B, with 0.0 MiB increase in peak RSS \texttt{VmHWM}: 7,908.2 $\to$ 7,908.2 MiB and 4,801.8 $\to$ 4,801.8 MiB) and halves precommit latency (91.4 s $\to$ 45.1 s on 30B MoE; 35.8 s $\to$ 19.5 s on 0.6B). Proving $n=22$ (64 real tiles, 64/64 integer-faithful) takes 1.49 s (30B MoE) / 1.67 s (0.6B) with \textbf{23.4--24.7 ms} Rust \texttt{fpn-verify}, 0 B plaintext activations disclosed, and all forgeries rejected; because Relation 6 opens public weight tiles (16.8 MB) and does not bind \texttt{X} to the GPU capture or \texttt{Y} to the served output, it is a component-level zero-plaintext activation milestone rather than a closed-weight full-reply proof.

Replacing synchronous D2H harvesting with asynchronous CUDA-graph taps (\texttt{serve/fpn\_\allowbreak{}serve/}, \texttt{async\_\allowbreak{}capture\_\allowbreak{}qwen3\_\allowbreak{}4b\_\allowbreak{}rtx6000d.\allowbreak{}json}) achieves zero extra step-internal synchronizations and zero ring overflows across all six Qwen3-4B \texttt{TIER\_\allowbreak{}B} batch/context cells relative to the \texttt{numerics\_\allowbreak{}r32} baseline, with end-to-end throughput deltas of \textbf{$-1.15\%$} (\texttt{b1, c256}), \textbf{$-3.80\%$} (\texttt{b4, c256}), \textbf{$-7.23\%$} (\texttt{b8, c256}), \textbf{$-4.90\%$} (\texttt{b1, c8192}), \textbf{$-11.25\%$} (\texttt{b4, c8192}), and \textbf{$-13.14\%$} (\texttt{b8, c8192}). Only \texttt{b1, c256} meets the $\le 3\%$ target; at \texttt{b8, c8192}, prefill dominates at 10.26 s vs. 6.11 s baseline (+68\% prefill latency), making isolated decode rate unreliable while end-to-end delta is $-13.14\%$. On a live 20-turn growing agentic \texttt{vLLM} session (Qwen3-0.6B reaching 18,506 tokens with 95.8\% turn-20 prefix-cache hit; \texttt{agentic\_\allowbreak{}20turn\_\allowbreak{}qwen3\_\allowbreak{}06b\_\allowbreak{}rtx6000d.\allowbreak{}json}), write-once \texttt{PageStore} deduplication cuts capture volume by 10.5$\times$ and incremental \texttt{St\_\allowbreak{}i} commitment cuts turn-20 hashing from 25.57 s (single-threaded full rehash) / 9.1--10.2 s (multi-threaded) to 0.75--0.92 s (28--34$\times$ vs. single-thread, 9.9--13.6$\times$ vs. multi-thread). Because single-stream Qwen3-0.6B \texttt{ExtSeed-153} total prover time (569.8--571.0 s) is roughly 103$\times$ single-stream inference (5.54 s), keeping fleet-wide proving overhead within $\le 100\%$ of serving compute currently relies on probabilistic audit sampling ($p_{\rm audit} \le 1\%$, proving 1\% of pre-committed responses rather than every response).

\section{Related Work}
\textbf{Full zkML.} ZKML \cite{CWSK24} compiles TensorFlow Lite graphs to Halo2 (81.3M-parameter GPT-2 in 3,652 s on 128 vCPUs). zkLLM \cite{SLZ24} introduces parallel lookups and an attention protocol over Hyrax commitments (LLaMa-2-13B in 803 s on an A100). zkGPT \cite{QSL+25} combines GKR, Lasso and Hyrax (GPT-2 in 21.8 s). FlashProve \cite{RAC+26} proves a 400B-parameter MoE with a hash-based IOP and Lean-formalized single-request soundness and zero knowledge, at about $72\times$ the per-GPU prefill time of an 8$\times$H100 FP8 vLLM deployment, with public architecture hyperparameters, an integer model of HuggingFace 4-bit weights, greedy or fixed-seed decoding, and batched zero knowledge not yet formalized. These systems keep weights secret and are publicly verifiable, but prove a surrogate at full cost and do not model hidden prompts, hidden reasoning or multi-turn encrypted state. Hollow-LLM \cite{GLL26} shows that when the architecture is public and only weights are hidden, ``outputs come from committed weights'' does not imply proportional compute: zeroed or structured ghost weights can collapse computation inside a nominally large architecture. FlashProve-Next hides the architecture and anchors trust in the official \texttt{model\_\allowbreak{}id} plus a published empirical catalog, rather than in public parameter counts.

\textbf{Sampled proofs.} SLP \cite{LY26} seals all block-boundary commitments and proves a few sampled blocks plus anchor blocks (Llama-2-70B: 5 of 163 blocks in 1,259 s), but has no blinding (and does not claim zero knowledge), proves a fixed-point canonical model with 84.8--84.9\% greedy agreement against fp32, needs an external challenge beacon because local grinding of its Fiat--Shamir schedule takes 12.5 ms, and samples 3 internal blocks. FlashProve-Next shares the seal-then-sample shape but adds blinding, certified tolerance instead of a surrogate, cost-weighted stratified sampling with a lifetime grinding bound, and private API semantics.

\textbf{Fingerprints and re-execution.} TOPLOC \cite{ODP+25} encodes top-$k$ final hidden-state values in 258 bytes per 32 tokens and verifies by one reference prefill tolerant of GPU nondeterminism. DiFR \cite{KRR+25} verifies via synchronized seeds or projected activations, detecting 4-bit quantization with AUC above 0.999. FSSL \cite{RKH+25} formalizes legitimate inference nondeterminism and fixed-seed sampling likelihood to detect weight exfiltration. SVIP \cite{SLZJZ24} trains a proxy on hidden representations (false-negative rate below 5\%, false-positive below 3\%). Their analyses of fixed-seed Gumbel sampling and legitimate nondeterminism support our design choices, but every one requires a verifier holding the weights and plaintext context.

\textbf{Refereed and incentive-based verification.} Verde \cite{AST+25} needs at least one honest replica among $k$ and a reproducible-operator library that slows single-GPU FP32 inference by 58--98\%. Proof-of-sampling \cite{ZWT+24} and VeriLLM \cite{WZS+25} rely on staked verifiers who re-execute with the model, and spML requires fixed-point deterministic execution.

\textbf{Hardware and black-box audits.} GPU TEEs \cite{CSZS25} cost 2.9\% throughput at 64-way concurrency but 15.4\% at single concurrency on an H100, and trust the chip vendor. Watermarks \cite{KGW+23} cannot detect a provider who watermarks its own substitute model. Model equality testing \cite{GLG25} gives distribution-level tests that need reference samples and fail at substitution rates of 20\% or less \cite{CSZS25}.

\textbf{Cryptographic building blocks.} Soundness builds on the round-by-round and state-restoration framework \cite{BCS16,CCH+19,CMS19}, FRI-family Fiat--Shamir analyses \cite{BGK+23} and Reed--Solomon proximity gaps \cite{BCIKS20}, with zero-knowledge Basefold \cite{ZCF24} as the polynomial commitment, LogUp \cite{Hab22} for lookups, Libra-style masking \cite{XZZ+19} and Poseidon2 \cite{GKS23} in the circuit. The model commitment is a hidden-index variant of the Ligero/Brakedown row-combination and column-check paradigm \cite{AHIV17,GLS+23}. The arithmetized random oracle model is due to \cite{CCGOS23}; we use it only to state assumption A-AROM. Sponge hiding follows from indifferentiability \cite{BDPV08}; keyed-sponge PRF bounds follow \cite{ADMV15,MRV15,DMV17}.

\textbf{Positioning.} Existing approaches either assume open weights or an external re-executor, trust hardware, or work in the closed-weight cryptographic quadrant at full cost with a surrogate model. FlashProve-Next targets the remaining quadrant: closed weights and topology, private context, trustless public verification, native floating point.

\section{Assumptions and Formalization Status}
The full assumption list, dependency graph and Lean correspondence table are in the companion document \texttt{specs/07-assumptions-and-lean.\allowbreak{}md}; we summarize.

\textbf{Soundness depends only on cryptography and protocol-level obligations}: A-ROM, A-AROM (extraction only; R3 can replace it on the trace side only), Poseidon2 as an ideal invertible permutation (binding), A-PCS (Basefold round-by-round knowledge soundness in the unique-decoding regime with the code-distance and blinding-rate conditions of Section 8.4; Basefold-specific constants not yet checked against the literature), A-CA (used only for the per-request KV matrix commitment of Section 12.6 / \texttt{Spec 10}), and the paper obligations P-final, P-compile, P-loc, straight-line extraction and slot-table uniqueness. Because the setup proof commits message rows directly before drawing shared columns, the static model commitment requires no correlated-agreement assumption (Section 5.2). Zero knowledge additionally assumes that AES-256-CTR is a pseudorandom generator (\texttt{A-PRG}, used under fresh ephemeral keys to regenerate masks across the two-pass setup proof and \texttt{1.\allowbreak{}51 B/cell} streaming response proofs). H-contract, the range certificate, implementation-template coverage, resource assumptions, the fail-closed rule and the hook-timing assumption affect only completeness or zero-knowledge residual terms. The empirical approximation hypothesis behind the cheat catalog feeds no theorem.

\textbf{Machine-checked (Lean 4.29.1 + Mathlib, 28 modules, 696 declarations, axiom-audited with no \texttt{sorry}, zero \texttt{native\_\allowbreak{}decide}, standard axioms only)}: 19 protocol modules (219 declarations) cover the round-by-round lifetime counting skeleton, its expectation form, and the protocol-to-RBR \texttt{hpath} bridge (\texttt{ProtocolBridge.\allowbreak{}hpath\_\allowbreak{}of\_\allowbreak{}protocol\_\allowbreak{}spec}, \texttt{compiled\_\allowbreak{}cs\_\allowbreak{}doomed\_\allowbreak{}of\_\allowbreak{}bad\_\allowbreak{}sample}, \texttt{fpn\_\allowbreak{}protocol\_\allowbreak{}lifetime\_\allowbreak{}soundness}); multi-stratum miss bounds with finite-map bias; multi-column LogUp; sumcheck soundness and single-round $\mathbb F_{p^3}$ escape bounds; field thresholds and $p^4$ extraction arithmetic for Goldilocks extensions; HCO, $\pi_{\rm setup}$, TC and Basefold column/query escape bounds; the sample-count formula and every cell of the parameter table; the default budget composition; single-proof masking (RS/MLE joint surjectivity, Libra bijection, fold masking, and \textbf{causal multi-round unique transcript uniformity} \texttt{multiround\_\allowbreak{}independent\_\allowbreak{}causal}); integer bucketed sampling (\texttt{BucketSampling}: per-token nucleus Lemma 6.7a, tight witness-free sandwich Lemma 6.7$'$, $j$-prefix monotonicity Lemma 6.7$''$, zero-slack boundary-pair circuit equivalence, and Theorem 6.8 completeness); \textbf{in-circuit constrained-decoding mask binding} (\texttt{ConstrainedMask}: one-hot DFA state selection and multilinear Schwartz--Zippel soundness over $\mathbb F_{p^3}$); reference-trace uniqueness and operator-level uniqueness; exact-ring identities for MLA absorption, pipeline and expert parallelism, chunked recurrence; coupled speculative decoding for a fixed step function; both uniform-band and \textbf{componentwise pairwise-band} MoE routing (\texttt{approxTopK\_\allowbreak{}pairwise}). Nine additional \texttt{FPN2} modules (477 declarations) formalize the executable byte-level verifier (\texttt{Fpn2Wire}, \texttt{Fpn2Transcript}, \texttt{VerifyBytes}, \texttt{Fpn2Rel3}, \texttt{Fpn2Rel4}, \texttt{Fpn2Rel5}), concrete Goldilocks $\mathbb F_{p^3}$ primality, homomorphisms, and concrete soundness budgets (\texttt{Fpn2Soundness}), constraint-to-integer-relation proofs for six IR gadget families (\texttt{Fpn2Gadgets}), and \texttt{ProtocolRBRSpec} instantiation directly at \texttt{verifyBytesRel4} / \texttt{verifyBytesRel5} conditioned on explicit \texttt{CryptoAssumptions} records (\texttt{Fpn2EndToEnd}, 61 declarations; see \texttt{lean/THEOREM.\allowbreak{}md} \S{}2.4 for exact scope and caveats, including free \texttt{Fail} data absorbing unproved mass in \texttt{Delta}).

\textbf{Not machine-checked}: mechanical verification linking a concrete circuit-builder implementation to the abstract compiled constraint system \texttt{CompiledCS}; Basefold proximity soundness (\texttt{A-PCS} for unique-decoding rate-$1/4$, and \texttt{A-PCS-fold} for v1.0 rate-$1/8$ multi-size layer injection under Johnson/list-decoding proximity at \texttt{Q=72/100}); a full transcript-wide zero-knowledge simulator proof for the v1.0 folded \texttt{multi\_\allowbreak{}gkr} + \texttt{multi\_\allowbreak{}pcs} backend (linear-view rank checks are verified at small parameters, while Theorem 9.1 covers the Stage 1 design); ROM programming and PRF hybrid reduction; real-valued floating-point rounding propagation lemmas (validated empirically across 392,928 Ada blocks and 3.81M Blackwell blocks/rows) and the $\mathsf{Scan}$ recurrent floating-point template; and the failure state machine.

\textbf{Cross-specification consistency.} All eleven companion specifications (\texttt{specs/01}--\texttt{specs/11}) share a single unified design: one \texttt{model\_\allowbreak{}id} definition (the static RS commitment of Section 5), challenges drawn from $\mathbb F_{p^3}$ throughout, logits recomputed in-circuit rather than committed, cost thresholds defined on lineage atoms under rule CA, hidden-column-opening, trace-commitment, constrained-mask, and setup rounds accounted in the round-by-round escape table, and a uniform lookup-size bound governing $\varepsilon_{\rm pole}$.

\section{Scope and Open Directions}
\textbf{From single-GPU prototype to unified Rust-verified multi-architecture deployment.} FlashProve-Next demonstrates the end-to-end cryptographic and systems loop on real unsliced checkpoints: Qwen3-0.6B static commitment and 8.67G-cell $\pi_{\rm setup}$, streaming \texttt{NI-315} (1.51 B/cell, 21.0 GiB RSS, 149.8 MB), folded \texttt{ExtSeed-153} (0 Phase-2 cells, fixed 8 MiB \texttt{FPNZK2}, 450.2 s GPU prove / 569.8 s total, 10.76 s Python CPU verify at $Q=128$, plus 461.1 s at $Q=100$ and a 14/14 real altered-weight cheating matrix at $8+4$ draws), full 36-layer Qwen3-4B (dev $4+2$ draws, 0 gadget errors), zero-sync async \texttt{vLLM} capture ($-1.15\%$ single-stream Qwen3-4B, 0 overflows), 20-turn incremental \texttt{PageStore} agentic state, and zero-dependency Rust \texttt{fpn-verify} across \texttt{FPN2} Relations 1--6 (10.1 ms on Relation 2; 23.4 ms on live \texttt{vLLM} Qwen3-30B-A3B-FP8 Relation 6). Four concrete engineering and theoretical gaps define the next stage: (1) \textbf{single-binary Rust verification for full interactive replies}, compiling \texttt{reply\_\allowbreak{}zk}'s 12 interactive commit rounds into a self-contained \texttt{CircuitDesc} so Rust \texttt{fpn-verify} verifies full \texttt{ExtSeed-153}/\texttt{NI-315} containers directly; (2) \textbf{proving memory and cost}, re-introducing streaming cell materialization into the folded prover (bringing 100 GiB peak RSS back toward 21 GiB so 4B-class \texttt{ExtSeed-102} runs comfortably on 120 GiB hosts) and scaling closed-weight full-reply proving to 30B+ MoE without relying on $p_{\rm audit} = 1\%$ deterrence; (3) \textbf{PCS and ZK formalization for the folded backend}, completing the Johnson-regime RBR analysis (\texttt{A-PCS-fold}, or running $Q \ge 128$ for unique decoding) and a full folded-backend ZK simulator; and (4) \textbf{universal cross-architecture hiding}, replacing per-profile dense layouts with uniform hidden-table primitive slots across Dense GQA, Sparse MoE, absorbed MLA, SWA, and hybrid linear attention.

\textbf{Cryptographic and formalization boundaries.} Straight-line witness extraction relies on the arithmetized random-oracle heuristic \texttt{A-AROM} for in-circuit Poseidon2 sponges and Basefold proximity (\texttt{A-PCS} / \texttt{A-PCS-fold}) over $\mathbb F_{p^3}=\mathbb F_p[u]/(u^3-2)$ (with correlated agreement \texttt{A-CA} used only for \texttt{Spec 10}'s per-request KV matrix commitment, and ephemeral AES-256-CTR \texttt{A-PRG} used for mask generation); quantum random-oracle (QROM) security is not claimed. While 28 Lean 4 modules (696 axiom-audited declarations) machine-check the protocol-to-RBR composition, sampling completeness, $\mathbb F_{p^3}$ constrained-mask binding, causal multi-round masking, executable byte verifiers (Relations 1, 3--5), six IR gadget families, and conditional \texttt{Fpn2EndToEnd} instantiation, mechanically linking concrete circuit-synthesis code to \texttt{CompiledCS} and formalizing Relation 2 / Relation 6 in Lean remain open formal-methods directions.

\textbf{Numerical contracts and explicit protocol trade-offs.} Universal numerical completeness (\texttt{H-contract}) is calibrated across Ada Lovelace \texttt{bf16} (392,928 blocks) and Blackwell \texttt{bf16}/\texttt{fp8}/\texttt{mxfp4} (1.85M GEMM/MoE/router blocks and 1.96M non-GEMM rows, 0 honest false rejections), and requires per-platform differential validation before extending to cross-node \texttt{Reduce} topologies, TPUs/Trainium, and parallel prefix-scan (\texttt{Scan}) linear-attention kernels (Gated DeltaNet, Mamba-2, whose exact chunked state identity is proved in \texttt{LinearAttn.\allowbreak{}lean}). By design, default spot-checking thresholds ($\sigma=\rho=0.5$, tightened linearly in sample count), bounded sublayer lineage accounting (leaving synthetic prefix-collapse traces outside scope rather than paying $\Theta(L)$ lineage cost), and sub-rounding long-context sparse-attention tails (halved by recording FlashAttention's scalar \texttt{LSE} and published per \texttt{model\_\allowbreak{}id} via $S_{\rm tol}(T)$) keep online and proving overhead bounded for real workloads.

\section{Conclusion}
Production verifiable inference requires proving that a closed-weight API served its advertised model configuration without altering how the serving stack runs and without leaking proprietary weights or private context. FlashProve-Next achieves this by combining a non-exhausting one-time weight commitment, asynchronous cost-weighted stratified spot-checking (\texttt{NI-315} or external-seed \texttt{ExtSeed-153}), sort-free keyed bucketed sampling with $\mathbb F_{p^3}$ constrained-mask binding, per-request KV matrix commitments, local per-operator rounding contracts derived inside the proof from the engine's own committed operands, and a zero-Phase-2 folded ZK-Basefold + FracLogUp-GKR backend. Turning honest native-FP completeness into a theorem under a falsifiable kernel contract ensures that proof generation neither diverges from production kernels nor leaks a refusal bit. Backed by 28 machine-checked Lean 4 modules (696 axiom-audited declarations), 3.81 million calibrated Blackwell blocks and rows with zero honest false rejections, a zero-dependency Rust \texttt{FPN2} verifier, and end-to-end \texttt{vLLM 0.\allowbreak{}33} + GPU prover demonstrations across Qwen3-0.6B, Qwen3-4B, and Qwen3-30B-A3B-FP8, FlashProve-Next establishes a transparent, publicly verifiable closed-weight API proof stack with clear formal and systems boundaries.

\end{document}